\documentclass[%
 reprint,
superscriptaddress,
showpacs,preprintnumbers,
nofootinbib,
 amsmath,amssymb,onecolumn,notitlepage,
 bm,
 aps,
]{revtex4-1}

\usepackage[dvipdfmx]{graphicx}
\usepackage{here}
\usepackage{amssymb}
\usepackage{comment}
\usepackage{physics}
\usepackage{epstopdf}
\usepackage{dcolumn}
\usepackage{booktabs}
\usepackage{bm}
\usepackage[dvipsnames]{xcolor}
\usepackage[caption=false]{subfig}
\usepackage{mhchem}

\usepackage{tensor}
\usepackage[dvipsnames]{xcolor}
\usepackage{hyperref}
\hypersetup{
colorlinks=true,
linkcolor=blue,
citecolor=BlueViolet,
urlcolor=Aquamarine
}

\newcommand{\deltaK}{\delta^\textsc{k}}
\newcommand{\cT}{\mathcal{T}}
\newcommand{\cH}{\mathcal{H}}

\usepackage[normalem]{ulem}

\usepackage[export]{adjustbox}

\newcommand{\mnras}{Mon. Not. R. Astron. Soc.}

\providecommand{\sorthelp}[1]{}

\begin{document}


\title{
Gravitational wave fossils in nonlinear regime:
halo tidal bias and intrinsic alignments from gravitational wave separate universe simulations
}

\newcommand{\IAS}{School of Natural Sciences, Institute for Advanced Study, 1 Einstein Drive, Princeton, NJ 08540, USA}

\author{Kazuyuki Akitsu}
\email{kakitsu@ias.edu}
\affiliation{\IAS}
\author{Yin Li}
\affiliation{Center for Computational Astrophysics
\& Center for Computational Mathematics, Flatiron Institute,
162 5th Avenue, New York, NY 10010, USA}
\author{Teppei Okumura}
\affiliation{Institute of Astronomy and Astrophysics, Academia Sinica, No. 1, Sec. 4, Roosevelt Road, Taipei 10617, Taiwan}
\affiliation{Kavli Institute for the Physics and Mathematics of the Universe (WPI), UTIAS, The University of Tokyo, 5-1-5 Kashiwanoha, Kashiwa, Chiba 277-8583, Japan}

\date{\today}
    
\begin{abstract}
We investigate impacts of long-wavelength gravitational waves (GWs) on nonlinear structure formation by utilizing the tidal separate universe simulations.
Based on the equivalence of a long-wavelength GW to a uniform tidal field in a local frame, 
we provide a way to incorporate a long-wavelength GW into the tidal separate universe simulation as an effective anisotropic expansion.
This methodology enables us to study effects of GWs on large-scale structure efficiently.
We measure the anisotropic imprint in the local power spectrum 
from the tidal separate universe simulations with GWs, which corresponds to the scalar-scalar-tensor bispectrum in squeezed limit or the so-called power spectrum response to GWs.
We also detect the halo tidal bias induced by GWs from the response of the halo-matter cross-power spectrum to GWs, as well as the linear shape bias (or the linear alignment coefficient) induced by GWs from the one-point function of the halo ellipticity.
In contrast to the case of the tidal field induced by scalar perturbations, we discover that the wavenumber dependence of the temporal evolution of GWs naturally causes these biases to be scale-dependent. 
We also find that this scale dependence is well approximated by the second-order density induced by the coupling between scalar and tensor perturbation.
This highlights that the structure formation, especially the process to determine the halo shape, is nonlocal in time.
Our findings lay the foundation for predicting the impact of GWs on large-scale structure.
\end{abstract}

\maketitle

\section{Introduction}
 
Gravitational waves (GWs) serve as vital means to observe the universe.
In particular, since long-wavelength GWs in the Gpc-Mpc range are unlikely to originate from astrophysical events, they are thought to have a cosmological origin (e.g., inflation~\cite{Starobinsky:1979ty,Sato:1980yn,Guth:1980zm,Linde:1981mu,Maggiore:1999vm}), making them  an important probe in cosmology.
In spite of the various experiments trying to detect the primordial $B$-mode signal in the polarization of the Cosmic Microwave Background (CMB)~\cite{WMAP:2010qai,Planck:2018jri,BICEP2:2018kqh,POLARBEAR:2022dxa,CMB-S4:2016ple}, which is one of the most powerful methods to hunt for long-wavelength cosmological GWs~\cite{Zaldarriaga:1996xe,Kamionkowski:1996ks}, such GWs have not been observed yet.

Compared to the numerous studies for the effect of GWs on the perturbations of the CMB, that on large-scale structure (LSS) of the universe has not received as much attention.
As large-scale structure is dominantly sourced by scalar perturbations,
there is a long history for the study on how the scalar perturbations have shaped LSS, including the linear and nonlinear perturbation theory~\cite{Dodelson:2003ft,Suto:1990wf,Makino:1991rp,Bernardeau:2001qr,Baumann:2010tm} and the $N$-body simulation~\cite{Springel:2005mi,Potter:2016ttn,Springel:2020plp,Angulo:2021kes}.
On the other hand, there are fewer studies of the effect of the tensor perturbations (GWs) on nonlinear structure formation, as we list below.

There are two types of effects of GWs on LSS observables: the dynamical effect and the projection effect.
The former refers to the effect of GWs on nonlinear structure formation itself, whereas the latter refers to the effect of GWs on the light path emitted from distant galaxies to us.
In other words, the dynamical effect would be observed by a comoving observer in a local frame, while the projection effect comes from the mapping of observables from the galaxy's local frame to our local frame (at the earth), 
which includes the Sachs-Wolfe effect and the gravitational lensing effect caused by GWs.
Both effects have been studied by means of the perturbation theory;
Refs.~\cite{Kaiser:1996wk,Dodelson:2003bv,Yoo:2009au,Jeong:2011as,Schmidt:2012nw,Jeong:2012nu} formulated the projection effect on the galaxy clustering and the galaxy shape (shear) by GWs, while
Ref~\cite{Schmidt:2012nw} pointed out that long-wavelength GWs can also contribute the intrinsic alignments of galaxy shapes and Refs.~\cite{Masui:2010cz,Dai:2013kra,Schmidt:2013gwa} computed the second order matter density contrast induced by the coupling between scalar perturbations and long-wavelength tensor perturbations (GWs).
For the projection effect the perturbative treatment would be adequate.
For the dynamical effect, however, the perturbation theory breaks down in nonlinear scales and thus the nonlinear nature of LSS requires $N$-body simulations with GWs to capture the fully nonlinear impact of GWs on structure formation.
Furthermore, given that the biased tracers of LSS such as halos are themselves nonlinear objects,
such simulations are necessary to understand their biases to GWs even in the linear regime.

An $N$-body simulation with GWs is generally challenging because the usual $N$-body simulations are based on Newtonian gravitational dynamics in an expanding background whereas GWs are a purely general-relativistic effect.
The most straightforward way to introduce GWs into $N$-body simulations is to develop a general-relativistic cosmological simulation~\cite{Adamek:2013wja,Adamek:2015eda,Adamek:2016zes}, though Refs.~\cite{Adamek:2015eda,Adamek:2016zes} only considered the second-order (induced) tensor perturbations by scalar perturbations.\footnote{
Technically the induced tensor perturbations computed in Refs.~\cite{Adamek:2015eda,Adamek:2016zes} involve non-GWs contributions, which are tensor modes but not propagating waves. See e.g., Ref.~\cite{Domenech:2020xin} for detailed discussion on this issue.}
Also, since GWs are much smaller than scalar perturbations, it is difficult to single out the effect of GWs on structure formation from that of the scalar perturbations in this sort of simulations.

In this paper, we utilize a separate universe approach to circumvent this issue on $N$-body simulations with GWs.
In the separate universe approach, the influence of a long-wavelength perturbation is absorbed into the cosmic expansion observed in the local frame, thereby the local expansion becomes different from the global one.
Accordingly, nonlinear structure formation in the local region responds to this difference in the background expansions.
Using this technique, the response to the long-wavelength perturbation can be accurately measured in $N$-body simulations~\cite{Sirko:2005uz,gnedin2011,Baldauf:2011bh,Li:2014sga,Wagner:2014aka,Lazeyras:2015lgp,Li:2015jsz,Baldauf:2015vio}.
Recently, the anisotropic extension of the separate universe simulation was developed in Refs.~\cite{Stucker:2020fhk,Masaki:2020drx,Akitsu:2020fpg}. 
They considered a long-wavelength \textit{tidal} perturbation sourced by the long-wavelength scalar tides $\propto \left(\partial_i\partial_j - \frac13 \deltaK_{ij}\right)\Phi$~\cite{Dai:2015jaa,Akitsu:2016leq,Akitsu:2017syq,Akitsu:2019avy,li2018,Taruya:2021jhg}, and the perturbation was absorbed into the local background, making the local cosmic expansion anisotropic.

Taking advantage of the equivalence of a long-wavelength GW to a long-wavelength tidal field in a local region, we apply this tidal separate universe simulation to measure the impact of GWs on structure formation.
In contrast to the scalar case, the time evolution of GWs is scale-dependent (or wavenumber-dependent).
As a result, when mimicking the long-wavelength GW as the local anisotropic expansion, the anisotropic expansion rate would be different depending on the wavenumber of GWs, and so is the response of large-scale structure, 
as considered for isotropic scale-dependent long-wavelength perturbation in Refs.~\cite{Hu:2016ssz,Chiang:2016vxa,Chiang:2017vuk}.
The purpose of this paper is to generalize the tidal separate universe simulation for GWs in order to study the scale-dependent responses for long-wavelength GWs of different wavelengths.

The remainder of this paper provides a way to implement long-wavelength GWs into the tidal separate universe simulation and presents imprints of GWs on large-scale structure measured from newly developed simulations.
After giving a brief review about the local coordinates in the presence of GWs and perturbative results in Sec.~\ref{sec:local_frame_and_2LPT}, we construct the tidal separate universe with GWs in Sec.~\ref{sec:tidal_SU}.
In Sec.~\ref{sec:power_response}, we measure the power spectrum responses for matter auto-, halo-matter cross-, and halo auto-power spectra, which is related to the scalar-scalar-tensor bispectrum in squeezed limit.
Sec.~\ref{sec:tidal_bias} and Sec.~\ref{sec:IA_GW} are devoted to the measurements of the halo tidal bias and linear alignment coefficient (or the linear shape bias) from GWs, respectively.
We discuss possible observables in large-scale structure to probe GWs in Sec.~\ref{sec:Discussion}.
Throughout this paper we adopt cosmological parameters consistent with Planck result: $\Omega_{\rm r0}=4.1577\times  10^{-5}$, $\Omega_{\rm m0}= 0.3089$, $\Omega_{\Lambda 0}=0.6911$, $H_0=67.74$~\cite{Planck:2015fie}.

\section{Local frame and the perturbative results}
\label{sec:local_frame_and_2LPT}

Here we first introduce the local coordinates in the presence of long-wavelength GWs. We then briefly summarize the derivation of the second-order density perturbations induced by the coupling between scalar and tensor perturbations (GWs), following Ref.~\cite{Schmidt:2013gwa}.
We employ the Lagrangian perturbation formalism, which can be straightforwardly used to construct the tidal separate universe in the next section.

\subsection{Long-wavelength gravitational waves in the conformal Fermi coordinates}
\label{sec:GWs_in_local_frame}

\begin{figure}[tb]
\centering
\includegraphics[width=\textwidth]{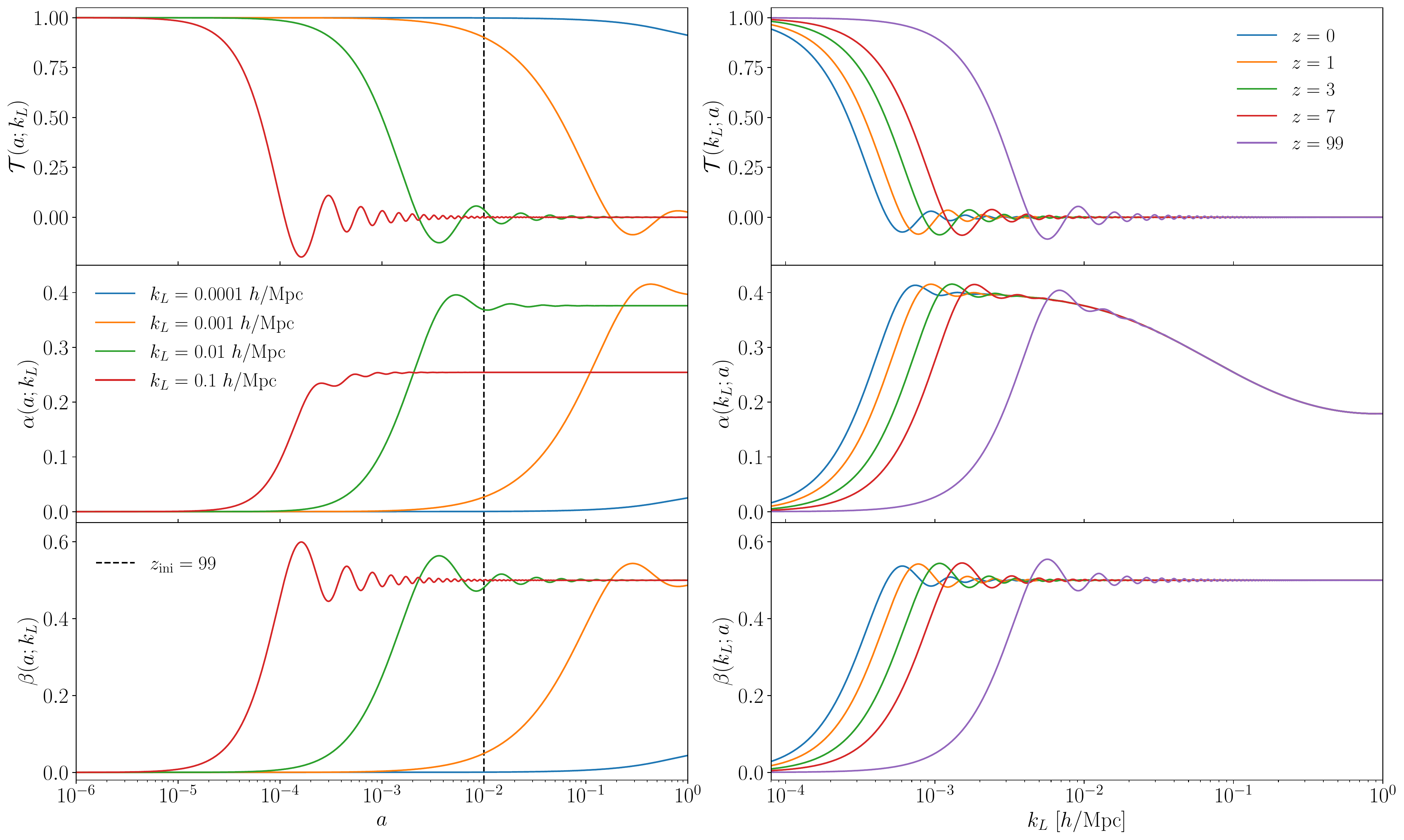}
\caption{
\textit{Left panels from top to bottom:} Transfer function of GWs $\cT(a;k_L)$, growth coefficient $\alpha(a;k_L)$ and dilation coefficient $\beta(a;k_L)$ in Eq.~\eqref{eq:2nd_density_final} as a function of scale factor $a$ for various wavenumbers $k_L$. 
\textit{Right panels from top to bottom:} 
Same as the left panels but as  a function of wavenumber of GWs, $k_L$, at various redshifts. 
Note that the functional shape of $\beta(a;k_L)$ is equivalent to that of $\cT(a;k_L)$ because $\beta(a;k_L) = -\frac{1}{2}\left[\cT(a;k_L)-1\right]$.
}
\label{fig:gw_trans_alpha_beta}
\end{figure}

In the cosmological context, GWs are defined as the trace-free transverse components in the perturbed FLRW metric,
\begin{align}
    ds^2 = a^2[-d\eta^2 + (\deltaK_{ij} +h_{ij})dx^idx^j ],
    \label{eq:FLRW_metric}
\end{align}
where $a$ is the scale factor, $\eta$ is the conformal time, $\deltaK_{ij}$ is Kronecker's delta, and $h_{ij}$ is GWs satisfying $\tensor{h}{^i_i} = 0$ and $\partial^i h_{ij}=0 $.
We use $k_L$ to denote the wavenumber of GWs in what follows since we focus on the long-wavelength GWs in this paper.
We introduce the transfer function of GWs, $\cT(\eta;k_L)$, through 
\begin{align}
    h_{ij}(\eta;{\bf k}_L) = \cT(\eta;k_L)h^{\rm ini}_{ij}({\bf k}_L),
\end{align}
where $h_{ij}^{\rm ini}(k_L)$ denotes the primordial value of GWs, i.e., $h_{ij}^{\rm ini}(k_L) = h_{ij}(0;k_L) $.
The transfer function $\cT(\eta;k_L)$ obeys
\begin{align}
    \cT''(\eta;k_L) + 2\cH\cT'(\eta;k_L) + k_L^2 \cT(\eta;k_L)= 0,
    \label{eq:gw_trans_diff_eq}
\end{align}
with $' = \dd/\dd\eta$, and $\cH = aH $ being the conformal Hubble paramter.
The top-left panel of Fig.~\ref{fig:gw_trans_alpha_beta} shows $\cT(\eta;k_L)$ as a function of the scale factor for various wavenumbers of GWs. Similarly, the top-right panel shows that as a function of the wavenumbers of GWs at various redshifts. 
For any $k_L$, $\cT(\eta;k_L)$ remains unity when $k_L\eta \ll 1$, which means that GWs are frozen before they enter the horizon.

For later convenience, we expand GWs by the helicity basis:
\begin{align}
    h_{ij}(\eta;{\bf k}_L) = \sum_{\lambda=\pm 2} h_{(\lambda)}(\eta;{\bf k}_L) e^{(\lambda)}_{ij}(\hat{k}_L),
\end{align}
where $e_{ij}^{(\pm 2)}\equiv e_i^{(\pm)}e_j^{(\pm)}$ and ${\bf e}^{(\pm)}\equiv ({\bf e}_1 \mp i {\bf e}_2)/\sqrt{2}$ with $\{{\bf k}_L, {\bf e}_1, {\bf e}_2 \}$ being an orthonormal set.
The helicity basis satisfies $e^{(\lambda)}_{ij}e^{ij}_{(-\lambda)} = 1$, $e^{(\lambda)}_{ij}e^{ij}_{(\lambda)} = 0$, and $h_{(\lambda)} = e^{ij}_{(-\lambda)}h_{ij}$.
The power spectrum of GWs for each helicity mode is defined as
\begin{align}
    \langle h_{(\lambda)}({\bf k}_L;\eta)h^*_{(\lambda)}({\bf k} '_L;\eta') \rangle = (2\pi)^3 \delta_{\rm D}^{(3)}( {\bf k}_L - {\bf k}'_L) P_{h_{(\lambda)}}(k_L;\eta,\eta'),
\end{align}
which is related to the primordial power spectrum $P_{h_{(\lambda)}}(k_L)$ through
\begin{align}
    P_{h_{(\lambda)}}(k_L;\eta,\eta') = \cT(\eta;k_L)\cT(\eta';k_L) P_{h_{(\lambda)}}(k_L).
\end{align}
For unpolarized GWs, the two power spectra has equal power, i.e., $P_{h_{(+2)}}(k) = P_{h_{(-2)}}(k) \equiv P_{h}(k)/2$ with $P_h(k)$ being the total power spectrum of GWs.
For chiral GWs, the chiral parameter $\chi(k)$ defined as
\begin{align}
    \chi(k) \equiv \frac{P_{(+2)}(k) - P_{(-2)}(k)}{P_h(k)}, 
    \label{eq:chi}
\end{align}
measures the degree of the parity-breaking in GWs.
The total power spectrum of GWs is often characterized via
\begin{align}
    \frac{k^3 P_h(k)}{2\pi^2}\equiv  r A_s \left(\frac{k}{k_*}\right)^{n_T}
\end{align}
with $r$ being the scalar-tensor ratio, $A_s$ being the amplitude for the primordial curvature perturbations at the pivot scale $k_*$, and $n_T$ being the tensor tilt.

To investigate the physical effects of long-wavelength GWs on scalar perturbations at smaller scales, consider a local region centered at the timelike geodesic of a comoving observer.
In this local patch we can construct the so-called conformal Fermi coordinates (CFC), which is an extension of the Fermi normal coordinates~\cite{Manasse:1963zz}, developed in Refs.~\cite{Schmidt:2013gwa,Dai:2015rda}. 
The metric of this coordinates, $g_{\mu\nu}^F$, takes the form of the FLRW metric along the central geodesic with leading-order corrections of ${\cal O}(x^2_F)$.
We are interested in the interaction between long-wavelength GWs and non-relativistic matter.
In this case $g_{00}^F$ encodes all the relevant impact from the long-wavelength perturbations because the dynamics of non-relativistic matter is solely determined by the usual Newtonian potential in $g_{00}^F$.
One can show that given the global coordinates Eq.~\eqref{eq:FLRW_metric}, $g_{00}^F$ is computed as 
\begin{align}
    g_{00}^F = -a^2\left[1+\tau_{ij}x_F^ix_F^j \right],
    \label{eq:CFC_metric}
\end{align}
with
\begin{align}
    \tau_{ij}(\eta; k_L) = -\frac12 \left[ a^{-1}\left(ah_{ij}'\right)' \right]
    = -\frac12 \left( h_{ij}'' + \cH h_{ij}' \right)
    \equiv T(\eta;k_L)h^{\rm ini}_{ij}
    \label{eq:CFC_tides}
\end{align}
representing the effective tidal field induced by the long-wavelength GWs in the local region.
The derivation is summarized in App.~\ref{app:CFC}.
It is worth noting that $\tau_{ij}$ takes effect only after GWs cross the horizon as implied by the time derivative
on $h_{ij}$.
Note that $x_F^i$ corresponds to the comoving distance, unlike the physical (proper) distance in the Fermi normal coordinates.
In what follows, we work in this frame and drop the subscript $F$ in $x_F^i$, namely, denote $x^i$ as $x_F^i$.

\subsection{Second order density induced by the interaction between GWs and scalar perturbations}
\label{subsec:2LPT_GWs}

The equation of motion of a matter particle in the local frame is 
\footnote{Strictly speaking, this is justified by considering the geodesic equation in the local frame.}
\begin{align}
    \frac{\dd^2 r_i}{\dd t^2} 
    = \frac{1}{a^2}\left[\frac{\dd^2}{\dd \eta^2}- \cH\frac{\dd}{\dd \eta} \right]r_i
    = -\frac{\partial}{\partial r_i}\left( \Phi_{\rm iso} + \phi \right),
    \label{eq:EoM_all}
\end{align}
where $r_i = a x_i$ and we split the gravitational potential into a background potential $\Phi_{\rm iso}$ and a peculiar potential $\phi$. The subscript  ``iso'' in $\Phi$ stands for the potential sourced only by the usual isotropic background, 
\begin{align}
    \Phi_{\rm iso} \equiv \frac23 \pi G\bar{\rho}_{\rm m}r^2 - \frac{\Lambda}{6}r^2,
    \label{eq:Phi_iso}
\end{align}
where $\bar{\rho}_{\rm m}$ is the mean density of matter.
The peculiar potential $\phi$ includes the potential sourced by local inhomogeneities $\phi_s$ as well as the effective tidal field sourced by the long-wave gravitational wave, $\phi = \phi_s + \frac12 \tau_{ij}x^ix^j$. In the next sections, the effective tidal potential is absorbed into the back ground potential, $\Phi = \Phi_{\rm iso}+\frac12 \tau_{ij}x^ix^j$ (see Eq.~(\ref{eq:background_potential}) below). 
The potential sourced by local inhomogeneities, $\phi_s$, satisfies the Poisson equation
\begin{align}
    \nabla_r^2\phi_s = 4\pi G \bar{\rho}_{\rm m} \delta = \frac32 \frac{\Omega_{\rm m}(\eta)\cH^2}{a^2}\delta,
    \label{eq:Poisson}
\end{align}
where $\delta = \rho / \bar\rho_\mathrm{m} - 1$ denotes the overdensity field.
We can also decompose the left-hand side of Eq.~\eqref{eq:EoM_all} into the background and peculiar parts, resulting in
\begin{align}
    \cH'x_i &= - \frac{\partial }{\partial x_i} \Phi_{\rm iso},
    \label{eq:Friedmann}
    \\
    x_i'' + \cH x_i' &= - \frac{\partial }{\partial x_i} \left(\phi_s + \frac12 \tau_{\ell m} x^\ell x^m\right).
    \label{eq:EoM_peculiar}
\end{align}
One can verify that Eq.~\eqref{eq:Friedmann} with Eq.~\eqref{eq:Phi_iso} gives the usual Friedmann equation.
Eq.~\eqref{eq:EoM_peculiar} can be regarded as the evolution equation for the displacement field $\Psi_i$, which relates Lagrangian position $q_i$ to Eulerian position $x_i$ via $x_i = q_i + \Psi_i$.
Before shell-crossing the Jacobian determinant of this mapping gives the overdensity as 
\begin{align}
    \delta = \left| \frac{\partial x_i}{\partial q_j} \right|^{-1} -1
    = \left| \deltaK_{ij} +  \frac{\partial \Psi_i}{\partial q_j} \right|^{-1} -1.
    \label{eq:Jacobian}
\end{align}
At linear order this reduces to $\delta^{(1)} =- \partial \Psi^{(1)}_i/\partial q_i$.
We split the linear displacement into the one sourced by pure scalar contributions and that sourced by $\tau_{ij}$,
each of which satisfies
\begin{align}
        \Psi^{(1)\prime\prime}_{s,i} + \cH \Psi^{(1)\prime}_{s,i} &= -\frac32 \Omega_m(\eta)\cH^2\frac{\partial_q^i}{\partial^2_q}\delta^{(1)},
        \\
        \Psi^{(1)\prime\prime}_{t,i} + \cH \Psi^{(1)\prime}_{t,i} &= -\frac12\partial_q^i\left[ \tau_{kl}q^k q^l\right],
\end{align}
where $\Psi^{(1)}_{s,i}$ and $\Psi^{(1)}_{t,i}$ denote the linear displacement caused by the scalar and tensor perturbations, respectively, and we have used Eq.~\eqref{eq:Poisson}
and the fact that $x_i$ can be replaced by $q_i$ at linear order.
The non-decaying solutions for $\Psi^{(1)}_{s,i}$ and $\Psi^{(1)}_{t,i}$ are
\begin{align}
    \Psi^{(1)}_{s,i}(\eta) &= -\frac{D(\eta)}{D(\eta_0)}\frac{\partial_q^i}{\partial_q^2} \delta^{(1)}(\eta_0),
    \\
    \Psi^{(1)}_{t,i}(\eta;k_L) &= \frac12\left[\cT(\eta;k_L)-1\right]h^{\rm ini}_{ij}(k_L) q^j
    \equiv -\beta(\eta;k_L) h_{ij}^{\rm ini}(k_L)q^j,
    \label{eq:disp_GWs}
\end{align}
where we have used the boundary condition $\lim_{\eta\to 0}\cT(\eta;k_L) = 1$ and have introduced the quantity $\beta(\eta;k_L)$ to characterize the linear displacement due to the tensor perturbations, and $D(\eta)$ represents the linear growth function, which follows
\begin{align}
    D''(\eta)+ \cH D'(\eta)- \frac32 \Omega_{\rm m}(\eta)\cH^2 D(\eta)=0.
    \label{eq:growth_func}
\end{align}
From this result one can confirm that GWs do not induce the linear density: $\delta^{(1)}_t = -\partial \Psi^{(1)}_{t,i}/\partial q_i \propto h_i^{\ i} =0$, as expected.

At second order, however, GWs do affect the density field as well as the displacement field. Expanding Eq.~\eqref{eq:Jacobian} up to second order leads to
\begin{align}
    \delta^{(1)}({\bf x}({\bf q}))+\delta^{(2)}_{st}({\bf x}({\bf q})) 
    =&
    \delta^{(1)}({\bf q})
     - \left.\frac{\partial \Psi^{(2)}_{st,i}}{\partial q_i}\right|_{{\bf q}}
     + \left.\frac{\partial \Psi^{(1)}_{s,i}}{\partial q_j}\frac{\partial \Psi^{(1)}_{t,j}}{\partial q_i}\right|_{{\bf q}}
     \nonumber\\
     =& \delta^{(1)}({\bf x}) - \Psi_{t,i}^{(1)}({\bf x})\partial_x^i \delta^{(1)}({\bf x})
     - \left.\frac{\partial \Psi^{(2)}_{st,i}}{\partial q_i}\right|_{{\bf q} ={\bf x}}
     + \left.\frac{\partial \Psi^{(1)}_{s,i}}{\partial q_j}\frac{\partial \Psi^{(1)}_{t,j}}{\partial q_i}\right|_{{\bf q} = {\bf x}},
     \label{eq:2nd_density}
\end{align}
where we focus on the second-order density arising from the coupling between the scalar and tensor modes, and $\delta_{st,i}^{(2)}$ and $\Psi_{st,i}^{(2)}$ respectively denote the second-order density and displacement induced by the coupling.
Namely, 
we omit the second-order contributions from the auto-coupling between scalars (irrelevant) or tensors (subdominant).
In deriving the first equality above, we have used the fact that $\Psi^{(1)}_{t,i}$ is divergence free.
Taking the divergence of Eq.~\eqref{eq:EoM_peculiar} with respect to ${\bf q}$ we obtain
\begin{align}
    \psi_{st}^{(2)\prime\prime} + {\cal H}\psi_{st}^{(2)\prime}
    = 
    -\nabla_x^2\phi_s 
    -\frac{\partial \Psi^{(1)}_{t,j}}{\partial q_i}\frac{\partial^2 \phi_s}{\partial x_i\partial x_j}
    -\frac{\partial \Psi^{(1)}_{s,j}}{\partial q_i}\tau_{ij},
\end{align}
where we have defined $\psi^{(2)}_{st} \equiv \partial \Psi_{st,i}^{(2)}/\partial q_i$ and used $\partial_q^i = \partial_x^i  + \partial_q^i\Psi_j \partial_q^j\simeq \partial_x^i  + \partial_q^i\Psi_j \partial_x^j$.
Rewriting $\phi_s$ in terms of $\delta$ by using Eq.~\eqref{eq:Poisson}
and using Eq.~\eqref{eq:2nd_density} yield
\begin{align}
    \psi_{st}^{(2)\prime\prime} + \cH\psi_{st}^{(2)\prime}
    = 
    \frac32 \Omega_{\rm m}\cH^2
    \left[ -\frac{\partial \Psi^{(2)}_{st,i}}{\partial q_i}
     + \frac{\partial \Psi^{(1)}_{s,i}}{\partial q_j}\frac{\partial \Psi^{(1)}_{t,j}}{\partial q_i}
    \right]
    -\frac32 \Omega_{\rm m}\cH^2\frac{\partial \Psi^{(1)}_{t,j}}{\partial q_i}\frac{\partial \Psi^{(1)}_{s,i}}{\partial q_j}
    -\frac{\partial \Psi^{(1)}_{s,j}}{\partial q_i}\tau_{ij}.
\end{align}
Finally the equation for $\psi_{st}^{(2)}$ is found to be
\begin{align}
    \psi_{st}^{(2)\prime\prime} + \cH\psi_{st}^{(2)\prime} - \frac32 \Omega_{\rm m} \cH^2\psi_{st}^{(2)} 
    =
    \left(\frac{\partial^i_q\partial^j_q}{\partial_q^2}\delta^{(1)}(\eta)\right)\tau_{ij}.
\end{align}
We can write the solution as 
\begin{align}
    \psi_{st}^{(2)}({\bf q}, \eta;k_L) = 
    D_{st}^{(2)}(\eta;k_L)
    \left(\frac{\partial^i_q\partial^j_q}{\partial_q^2}\delta^{(1)}({\bf q}, \eta_0)\right) h_{ij}^{\rm ini}(k_L),
\end{align}
where the time-dependent part of $ \psi_{st}^{(2)}({\bf q}, \eta;k_L)$, which we write $D_{st}^{(2)}(\eta;k_L)$, satisfies
\begin{align}
    D_{st}^{(2)\prime\prime}(\eta;k_L) + \cH D_{st}^{(2)\prime}(\eta;k_L) - \frac32 \Omega_{\rm m}(\eta)\cH^2 D_{st}^{(2)}(\eta;k_L)
    = \frac{D(\eta)}{D(\eta_0)} \cdot\left[ -\frac{1}{2a}\left(a\cT'(\eta;k_L)\right)'\right].
    \label{eq:Dt2_diff_eq}
\end{align}
We can solve this equation numerically given the cosmological parameters but here we derive the analytic solution assuming the matter domination for the later convenience.
In the matter-dominated era where we have $\Omega_{\rm m}=1$, $\cH=2/\eta$ and $D(\eta)\propto a \propto \eta^2$, we can write the solution using Green's function:
\begin{align}
    D_{st}^{(2)}(\eta;k_L)
    = & \frac{1}{D(\eta_0)}
    \int_0^\eta \dd\tilde{\eta}~ \frac15  \left[\frac{\eta^2}{\tilde{\eta}} - \frac{\tilde{\eta}^4}{\eta^3} \right]
    D(\tilde{\eta})\left[-\frac{1}{2a}\left[a\cT'(\tilde{\eta};k_L)\right]'\right]
    \nonumber\\
    = & \frac{1}{5}\frac{D(\eta)}{D(\eta_0)}
    \left[\beta(\eta) + 4\int_0^\eta \dd\tilde{\eta}\left(\frac{\tilde{\eta}}{\eta}\right)^5\beta'(\tilde{\eta};k_L)\right].
\end{align}

Finally we find the second order density induced by the interaction between the scalar and tensor perturbation as 
\begin{align}
    \delta^{(2)}_{st}({\bf x},\eta;k_L) 
    =& h_{ij}^{\rm ini}\left[\beta(\eta;k_L)x^j\partial_x^i \delta^{(1)}({\bf x},\eta)
    - D_{st}^{(2)}(\eta;k_L)
    \left(\frac{\partial^i_x\partial^j_x}{\partial_x^2}\delta^{(1)}({\bf x},\eta_0)\right)
    +\beta(\eta;k_L)\left(\frac{\partial^i_x\partial^j_x}{\partial_x^2}\delta^{(1)}({\bf x},\eta)\right)\right]
    \nonumber\\
    =&h_{ij}^{\rm ini}\left[\alpha(\eta;k_L)
    \frac{\partial^i_x\partial^j_x}{\partial_x^2}
    +\beta(\eta;k_L)x^j\partial_x^i
    \right]\delta^{(1)}({\bf x}, \eta),
    \label{eq:2nd_density_final}
\end{align}
where $\beta$ is defined in Eq.~(\ref{eq:disp_GWs}) and $\alpha$ is defined as
\begin{align}
    \alpha(\eta;k_L)\equiv &
    - \frac{D(\eta_0)}{D(\eta)}D_{st}^{(2)}(\eta;k_L) + \beta(\eta;k_L)
    \nonumber\\
    = &\frac45\left(\beta(\eta;k_L) - \int_0^\eta \dd\tilde{\eta}\left(\frac{\tilde{\eta}}{\eta}\right)^5\beta'(\tilde{\eta};k_L) \right).
    \label{eq:def_alpha}
\end{align}
Eq.~\eqref{eq:2nd_density_final} holds not only during the matter domination but also the entire history of the universe as long as we solve Eq.~\eqref{eq:Dt2_diff_eq}
numerically,
whereas the second equality of Eq.~\eqref{eq:def_alpha} holds only for the matter domination.
The first term in the last line $(\propto \alpha)$ represents the changes in the short-mode amplitude, called the growth effect, by the coupling between the scalar tidal field and long-wavelength GWs. The second term in the last line $(\propto \beta)$ represents the coordinate shift induced by the long-wavelength GWs, known as the dilation effect.
The power spectrum of the density field in the presence of the long-wavelength GWs then becomes
\begin{align}
    P_{\rm mm}({\bf k_S},\eta|h_{ij}(k_L))
    = P_{\rm lin}(k_S)\left[1 + \hat{k}_S^i \hat{k}_S^j h_{ij}^{\rm ini}(k_L)\left(2\alpha(\eta;k_L) - \beta(\eta;k_L)\frac{\partial \ln{P_{\rm lin}(k_S,\eta)}}{\partial \ln{k_S}}\right) \right].
    \label{eq:matter_power_GWs_PT}
\end{align}

We show $\cT(k_L;\eta)$, $\alpha(k_L;\eta)$ and $\beta(k_L;\eta)$ as a function of the scale factor for various wavenumbers of GWs (the left panels) and as a function of the wavemunber of GWs at various redshifts at the right and left panels of Fig.\ref{fig:gw_trans_alpha_beta}, respectively.
To plot these functions, we numerically integrated Eqs.~\eqref{eq:gw_trans_diff_eq} and \eqref{eq:Dt2_diff_eq} without assuming the matter domination.
In the limit of $k_L\eta\to 0$, 
it is obvious that there is no physical effect from GWs, i.e., $\alpha(\eta;k_L)\to 0$ and $\beta(\eta;k_L)\to 0$, since GWs are frozen on the super-horizon scales.
On the other hand, taking the limit $k_L\eta \gg 1$ is more interesting because in this limit $\cT(\eta;k_L)\to 0$ but $\alpha(\eta;k_L)$ and $\beta(\eta;k_L)$ do not vanish.
In other words,
even long after GWs have decayed away,
their impact on the growth and displacement remains,
which is sometimes called the ``fossil'' effect \cite{Masui:2010cz}.

\section{Tidal separate universe with gravitational waves}
\label{sec:tidal_SU}
In this section, we describe how to incorporate long-wavelength GWs into the simulation background with the help of the tidal separate universe simulation technique developed in Ref.~\cite{Akitsu:2020fpg} (see also Refs.~\cite{Stucker:2020fhk,Masaki:2020drx}).  
We focus on differences arising from GWs and refer the readers to Ref.~\cite{Akitsu:2020fpg} about the details of the implementation.
In this section we do not employ Einstein's summation convention to avoid confusions.

\subsection{Anisotropic background}
\label{subsec:TSU_background}

In the tidal separate universe simulation we introduce anisotropic scale factors by absorbing the long-wavelength tidal perturbations into the background.
In general, reflecting that the tidal perturbations are expressed by the $3\times 3$ symmetric matrix $\tau_{ij}$ the anisotropic scale factors are also written by the $3\times 3$ symmetric matrix $a_{ij}$, which relates the physical coordinate $r_i$ to the comoving coordinate $x_i$ as $r_i = \sum_j a_{ij} x_j$.
However, we can always rotate the simulation coordinates to align with the eigenvectors of $\tau_{ij}$,
leaving only the diagonal components non-zero so that $\tau_{ij} = \tau_{i}\deltaK_{ij}$ and $a_{ij} = a_{i}\deltaK_{ij}$.
Now we have different scale factors for each axis and characterize these differences by $\Delta_i$ defined via
\begin{align}
    a_i = a (1+ \Delta_i),
\end{align}
with $a$ being the global scale factor.
While the equation of motion remains the same as in Sec.~\ref{subsec:2LPT_GWs},
now we want to absorb the long-wavelength effective tidal potential induced by GWs into the background potential so that
\begin{align}
    \Phi = \Phi_{\rm iso} + \frac12 \sum_i\tau_i x_i^2  = \frac23 \pi G\bar{\rho}_{\rm m}r^2 - \frac{\Lambda}{6}r^2 
            + \frac12\sum_i \tau_i x_i^2. \label{eq:background_potential}
\end{align}
The background and peculiar equations in an anisotropic background become
\begin{align}
    \frac{1}{a^2}\left[ a_i'' - \cH a_i' \right] x_i 
    = -\frac{1}{a_i}\frac{\partial}{\partial x_i} \Phi,
    \label{eq:aniso_backgroud}
    \\
    \frac{1}{a^2}\left[ a_i x_i'' + 2a_i'x_i' - \cH a_i x_i \right]  
    = -\frac{1}{a_i}\frac{\partial}{\partial x_i} \phi_s.
    \label{eq:aniso_peculiar}
\end{align}
Subtracting the isotropic background, which is determined by Eq.~\eqref{eq:Friedmann}, from Eq.~\eqref{eq:aniso_backgroud} and linearizing it in $\Delta_i$ we find
\begin{align}
    \Delta''_i(\eta;k_L) + \cH\Delta_i'(\eta;k_L) 
    = \frac12 a^{-1}(\eta) \left[ a(\eta)  h_i'(\eta;k_L)\right]'.
    \label{eq:Delta_i}
\end{align}
Note that anisotropic scale factors depend on the wavenumber of GWs as a consequence of the wavenumber dependence of the transfer function of GWs, $\cT(\eta;k_L)$, while they do not for the case of scalar large-scale tidal field where the linear growth of scalar perturbations is independent of their wavenumbers, as characterized by $D(\eta)$.
Note also that the source term for the anisotrpic scale factors on the right-hand side is non-zero only when $h_i'(\eta;k_L)$ does not vanish; in other words long-wavelength GWs induces the anisotropic scale factors only after GWs enters the horizon as expected.
In fact, integrating Eq.~\eqref{eq:Delta_i} twice yields
\begin{align}
    \Delta_i(\eta;k_L) = \frac{h^{\rm ini}_i(k_L)}{2} \left[\cT(\eta;k_L) -1 \right]
    = - h^{\rm ini}_i(k_L) \beta(\eta;k_L),
    \label{eq:Delta_i_beta}
\end{align}
which goes to zero when $\eta\to 0$ because $\lim_{\eta\to 0}\cT(\eta;k_L)=1$.

The appearance of the function $\beta(\eta;k_L)$ defined in Eq.~\eqref{eq:disp_GWs} is expected and 
this result can be understood in a more intuitive way.
What is specifically done in the separate universe construction is to absorb the displacement caused by the long-wavelength perturbation into the background expansion while keeping the physical distance unchanged.
In other words, we introduce the local scale factor $a_i$ to satisfy
\begin{align}
    a_i x_i = a (x_i + \Psi_i^{\rm long}),
\end{align}
which implies $\Delta_i x_i  = \Psi_i^{\rm long}$.
Given the displacement caused by long-wavelength GWs in Eq.~\eqref{eq:disp_GWs}, this immediately leads to $\Delta_i(\eta;k_L) =- h_i^{\rm ini}(k_L)\beta(\eta;k_L)$.

We note that this matching only works for the non-relativistic matter. 
In other words, the effect of long-wavelength GWs can be captured by the anisotropic expansion only when we focus on non-relativistic particles that do not care about $g_{0i}$ and $g_{ij}$ components in the metric.
For example, the method presented here is not useful in order to study the impact of long-wavelentgh GWs on the radiation perturbations.
However, this treatment is sufficient to study the influence on dark matter particles that we are interested in and consistent with the usual Newtonian $N$-body method.

\subsection{Initial conditions}
\label{subsec:TSU_ICs}

The background anisotropy induces a correction to the 2LPT solution in the isotropic background as discussed in Ref.\cite{Akitsu:2020fpg}.
The correction depends on $\Delta_i(\eta)$, which is different for the scalar tidal field and GWs.
Here we derive this correction induced by the background anisotropy governed by long-wavelength GWs.

The equation for the displacement in the anisotropic background can be obtained by combining Eq.~\eqref{eq:aniso_peculiar} with Eq.~\eqref{eq:Poisson} as 
\begin{align}
    \sum_{ij} \left| \frac{\partial {\bf x}}{\partial {\bf q}} \right|
    \left[ \delta_{ij} + \Psi_{i,j} \right]^{-1}
    \left[ \Psi_{i,j}'' + (\cH+2\Delta_i')\Psi_{i,j}'\right]
    = \frac32 \Omega_{\rm m} \cH^2 \left(\left| \frac{\partial {\bf x}}{\partial {\bf q}} \right| - 1\right),
\end{align}
where we adopt the notation $\Psi_{i,j} \equiv \partial_q^j \Psi_{i}$ in this subsection.
Taking $\Delta_i=0$ results in the usual master equation for the LPT.
We introduce the correction $\epsilon^{(1)}_i$ of order ${\cal O}(\delta^{(1)}\Delta_i)$ with $\delta^{(1)}$ being short-wavelength modes in the simulations:
\begin{align}
    \Psi_i = \Psi_i^{(1)} + \Psi_i^{(2)} + \epsilon^{(1)}_i,
\end{align}
where $\Psi_i^{(1)} + \Psi_i^{(2)}$ is the usual 2LPT solution in the isotropic background. 
In the following we also introduce the potentials such that $\Psi_i^{(1)} = - \psi^{(1)}_{,i}$ and $\epsilon_i^{(1)} = - \epsilon^{(1)}_{,i}$ for the convenience.
The equation for $\epsilon^{(1)}$ can be found as 
\begin{align}
    \sum_i\epsilon_{,ii}^{(1)\prime\prime} + \cH\sum_i\epsilon_{,ii}^{(1)\prime} - \frac32 \Omega_{\rm m}\cH^2 \sum_i\epsilon_{,ii}^{(1)}
    = - 2\sum_i\Delta_i' \psi_{,ii}^{(1)\prime}.
\end{align}
Going to Fourier space and decomposing $\epsilon^{(1)}({\bf k})$ as $\epsilon^{(1)}({\bf k}) = \sum_i \hat{k}_i^2 \varepsilon_i^{(1)}({\bf k})$, this equation can be rewritten as
\begin{align}
    \varepsilon_i^{(1)\prime\prime} + \cH\varepsilon_i^{(1)\prime} - \frac32 \Omega_{\rm m}\cH^2 \varepsilon_i^{(1)}
    = - 2\Delta_i' \psi^{(1)\prime}.
\end{align}
Notice that $\varepsilon_i^{(1)}$ is different from $\epsilon_i^{(1)}$.
We can derive the matter-dominated solution for $\varepsilon_i^{(1)}$ for initial-condition generation. 
Since Green's function for this equation is the same as Eq.~\eqref{eq:Dt2_diff_eq} the solution is 
\begin{align}
     \varepsilon_i^{(1)}(\eta)
     =& \int_0^{\eta} \dd\tilde{\eta}~ \frac15 \left[\frac{\eta^2}{\tilde{\eta}} - \frac{\tilde{\eta}^4}{\eta^3} \right]
     \cdot\left(-2\Delta_i'(\tilde{\eta}) \psi^{(1)\prime}(\tilde{\eta})\right)
     \nonumber\\
     =&-\frac25 \frac{\psi^{(1)}(\eta)}{D(\eta)} \frac{\Delta_i(\eta)}{\beta(\eta)}
     \int_0^{\eta} \dd\tilde{\eta}~ \left[\frac{\eta^2}{\tilde{\eta}} - \frac{\tilde{\eta}^4}{\eta^3} \right]
     \cdot D'(\tilde{\eta}) \beta'(\tilde{\eta})
     \nonumber\\
     =&-\frac45 \psi^{(1)}(\eta) h_i^{\rm ini}
      \left[\beta(\eta) - \int_0^{\eta} \dd\tilde{\eta}~\left(\frac{\tilde{\eta}}{\eta}\right)^5\beta'(\tilde{\eta}) \right]
     \nonumber\\
     =& \psi^{(1)}(\eta) h_i^{\rm ini} \alpha(\eta),
\end{align}
where we have used $D(\eta) \propto \eta^2$ and $\Delta_i(\eta) = h_i^{\rm ini} \beta(\eta)$ and $\alpha(\eta)$ is introduced in Eq.~\eqref{eq:2nd_density_final}.
Thus
\begin{align}
    \epsilon^{(1)}(\eta) = \psi^{(1)}(\eta) \alpha(\eta)\sum_i h_i^{\rm ini}\hat{k}_i^2.
    \label{eq:epsilon}
\end{align}
This implies that the linear growth function has a direction-dependent modulation:
\begin{align}
    D(\eta,{\bf k}) = D(\eta)\left[1 + \alpha(\eta;k_L) \sum_i h_i^{\rm ini} \hat{k}_i^2 \right],
\end{align}
which is consistent with Eq.~\eqref{eq:2nd_density_final}.
Using Eq.~\eqref{eq:epsilon}, the correction to the velocity can be computed as 
\begin{align}
    \epsilon_{i}^{(1)\prime}
    =& \left( f_\alpha + f_1 \right) \cH\epsilon_{i}^{(1)},
\end{align}
with $f_\alpha \equiv \dd \ln \alpha / \dd \ln a$ and $f_1 \equiv \dd \ln D / \dd \ln a$.
We implement these modifications in \texttt{2LPTIC}~\cite{Crocce:2006ve} and generate the initial conditions at $z_i = 99$.

\subsection{Simulations}
\label{subsec:TSU_details}

\begin{figure}[tb]
\centering
\includegraphics[width=0.8\textwidth]{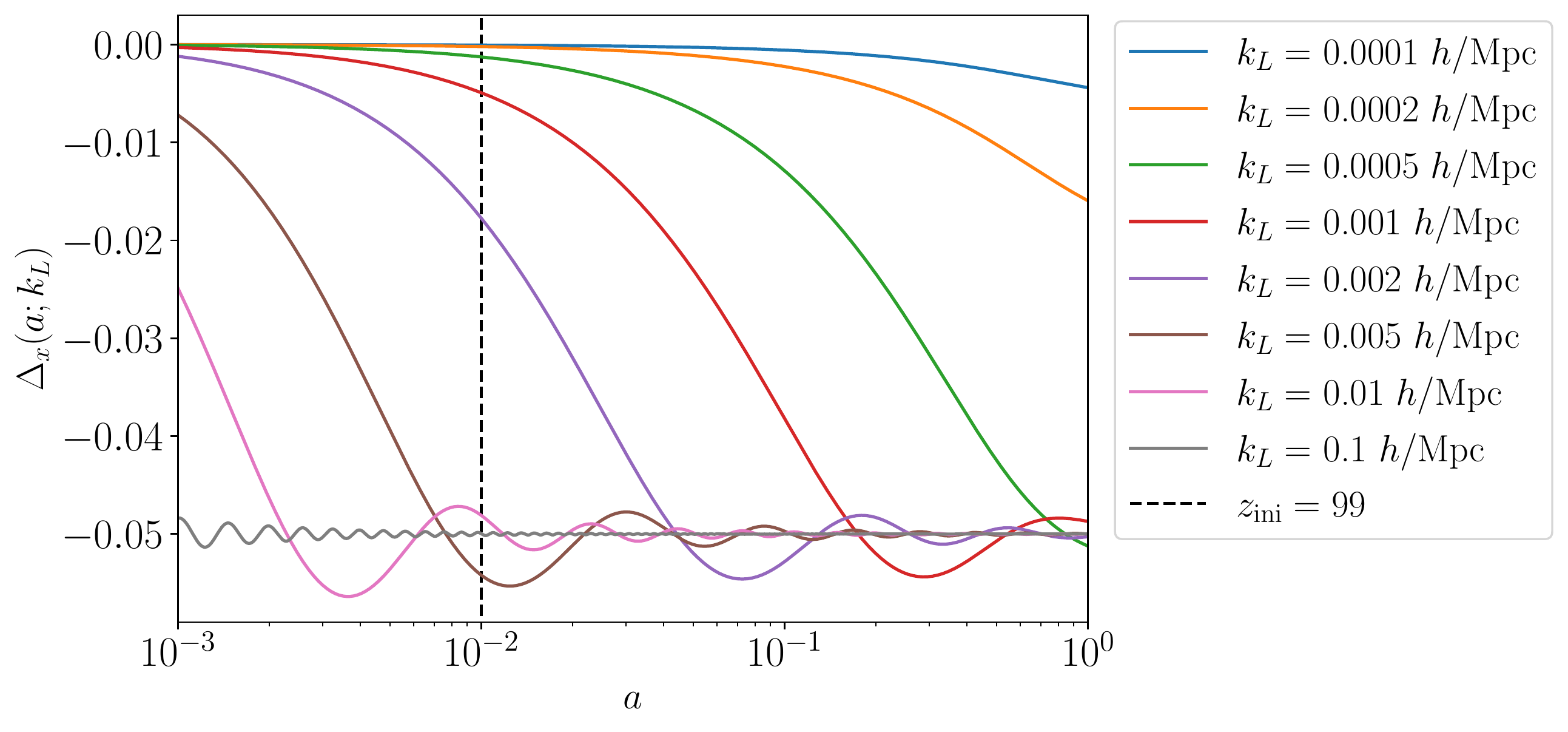}
\caption{
Fractional anisotropic scale factor in the $x$-axis $\Delta_x(a;k_L)= a_x(a;k_L)/a-1$ for various wavenumbers of GWs $k_L$ as a function of the scale factor $a$ in the case of $h^{\rm ini}_{ij} = {\rm diag}(0.1,-0.1,0)$.
The vertical dashed line represents the starting redshift $z_{\rm ini}=99$.
}
\label{fig:Delta_x}
\end{figure}

We perform $N$-body simulations in the tidal backgrounds with $1024^3$ particles in $500~\mathrm{Mpc}/h$ boxes.
The details of modifications of the $N$-body code based on \texttt{Gadget-2}~\cite{Springel:2005mi} in the tidal background is described in Ref.~\cite{Akitsu:2020fpg}.
One important additional modification to the code was made in the drift operator, which is discussed in Appendix~\ref{app:drift}.

After rotating $\tau_{ij}$ to align its eigenvectors with the simulation axis, the remaining degrees of freedom can be completely characterized by two parameters, which we can parametrize as $\tau_{\rm e} \equiv -(\tau_1 - \tau_2)/2$ and $\tau_{\rm p} = -\tau_3 + (\tau_1+\tau_2)/2$.
Taking into account the transverse condition of GWs, we cannot consider $\tau_{\rm p}$-type tides for the tidal separate universe with GWs unlike the case of the scalar tidal field in Ref.~\cite{Akitsu:2020fpg}.
Hence in this paper we only consider $\tau_{\rm e}$-type tides for the background anisotropy.
In other words, we consider GWs propagating along $z$ direction with $+$ mode polarization:
\begin{align}
    h^{\rm ini}_{ij} = 
    \begin{pmatrix}
    \pm\epsilon& 0& 0 \\
    0& \mp\epsilon& 0\\
    0& 0& 0
    \end{pmatrix},
    \label{eq:h_taue}
\end{align}
where we choose $\epsilon=0.1$. 
Since we aim to measure the \textit{response} of large-scale structure to the long-wavelength GWs, the results should not be dependent on the choice of the direction of the propagation and the polarization basis.

In order to investigate effects of GWs over a wide range of wavenumbers, we run tidal separate universe simulations with various wavenumbers of GWs:
\begin{align}
    k_L = \left\{ 0.0001,\, 0.0002,\, 0.0005,\,
    0.001,\, 0.002,\, 0.005,\,
    0.01,\, 0.02,\, 0.05,\,
    0.1,\, 0.2 \right\} \, [h/{\rm Mpc}].
\end{align}
These different wavenumbers of GWs give rise to the different time evolution of the local anisotropic scale factors, through which long-wavelength GWs affect the simulated small-scale structure formation.
Some examples of the time evolution of $\Delta_x$ are shown in Fig.~\ref{fig:Delta_x}.
Given the box size of $500~{\rm Mpc}/h$, some wavenumbers are larger than the fundamental mode in the simulation: $k_{\rm F}=0.013~h/{\rm Mpc}$.
For such larger wavemubers, we cannot neglect the curvature of the long-wavelength modes and treat GW as a unifiom tidal field over the whole simulation box and the approximation is violated.
Still, these GWs are longer-mode than the halo formation scale so GW can be seen as a uniform tide in that local region.
Therefore we can study the impact of GWs on halos, in particular, the response of the halo shape as discussed in Sec.~\ref{sec:IA_GW}, where we come back to this issue again.

For references, we also run fiducial simulations with the isotropic background, and tidal separate universe simulations induced by the scalar $\tau_{\rm e}$-type tidal field. These reference simulations share the same parameters (including random seeds) as those used in the GW tidal separate universe simulations. For each type of simulations (fiducial and tidal separate universe with GWs and scalar tides), we run four realizations, amounting to 100 simulations in total.

We use the \texttt{AHF} code~\cite{Knollmann:2009pb} to identify dark matter halos in the simulations by 
spherical overdensity (SO) regions 200 times as dense as the mean matter density.
We need to identify SO halos in the \emph{global} coordinates on the \emph{isotropic} background, i.e., $a_i x_i / a$, while \texttt{AHF} by default uses the local simulation coordinates $x_i$ on the \textit{anisotropic} background.
To this end, we modify the \texttt{AHF} code to rescale the coordinate correspondingly in distance computations.

\section{Anisotropic power spectrum response: tensor fossils in nonlinear regime}
\label{sec:power_response}

In this section, as a first example of imprints of GWs on large-scale structure, we present the anisotropic impact on the matter auto-, matter-halo cross-, and halo auto-power spectra by long-wavelength GWs measured from our $N$-body simulations.

\subsection{Growth-dilation decomposition}
\label{subsec:growth_dilation}

As we derived in Sec.~\ref{subsec:2LPT_GWs}, long-wavelength GWs leave the anisotropic imprint in the matter power spectrum in a given realization of GWs through the nonlinear interaction of tidal fields.
This tidal response consists of two different contributions as in Eq.~\eqref{eq:matter_power_GWs_PT}; the term proportional to $\alpha$ that modulates the amplitude of the power spectrum and the term proportional to $\beta$ that modulates the scales.
In terms of the tidal separate universe introduced in Sec.~\ref{subsec:TSU_background},
the former known as the growth effect describes the changes in the amplitude of short-mode fluctuations measured in the local anisotropic background,
while the latter known as the dilation effect stems from the anisotropic expansion of the local background with respect to the global one.
Specifically, the \textit{response} of the power spectrum to long-wavelength GWs can be decomposed as
\begin{align}
    \left.\frac{\dd{\ln P_{\cal G}}}{\dd{h^{\rm ini}_{ij}}}\right|_{{\bf k}_{\cal G}} =
    \left.\frac{\dd{\ln P_{\cal L}}}{\dd{h^{\rm ini}_{ij}}}\right|_{{\bf k}_{\cal G}}
    =&
    \left.\frac{\partial\ln P_{\cal L}}{\partial h^{\rm ini}_{ij}}\right|_{{\bf k}_{\cal L}}
    +\left.\frac{\partial\ln P_{\cal L}}{\partial \ln k_{{\cal L},i'}}\right|_{h_{ij}^{\rm ini}}
    \left.\frac{\dd{\ln k_{{\cal L},i'}}}{\partial h^{\rm ini}_{ij}} \right|_{{\bf k}_{\cal G}}
    \nonumber\\
    \equiv&\,
    \hat{k}_i\hat{k}_j
    \left[
    R^{\rm GW}_{\rm growth}(k;k_L)
    +R^{\rm GW}_{\rm dilation}(k;k_L)
    \right],
    \label{eq:def_response_GW}
\end{align}
where in the first line the power spectra defined with respect to the global isotropic background and the local anisotropic background are denoted by $P_{\cal G}$ and $P_{\cal L}$, respectively, and correspondingly the wavenumbers by ${\bf k}_{\cal G}$ and ${\bf k}_{\cal L}$.
The physical scale should be unchanged in these two coordinates, $a_i x_{{\cal L},i} = a x_{{\cal G},i}$, which implies $k_{{\cal L},i} = k_{{\cal G},i}(1+\Delta_i)$.
In the first equality we have used that the variances must be conserved in the coordinate transformation: $P_{\cal G}\, \dd^3 {\bf k}_{\cal G} = P_{\cal L}\,\dd^3 {\bf k}_{\cal L}$, together with $|\dd^3 {\bf k}_{\cal L}/\dd^3 {\bf k}_{\cal G} |= \prod_{i=1}^3 (1+\Delta_i)=1$ at leading order.
In the second line, We relabel both $k_{\cal G}$ and $k_{\cal L}$ as $k$,
since the responses $R^{\rm GW}_{\rm growth}$ and $R^{\rm GW}_{\rm dilation}$ are already first order in $h^{\rm ini}_{ij}$ (or $\Delta_i$) and hence here we do not need to distinguish $k_{\cal G}$ and $k_{\cal L}$.
Notice that we distinguish $k_L$ from $k_{\cal L}$; the former represents the wavenumber of GWs.
Notice also that we have defined the response with respect to the $h_{\rm ini}$, not $h(z)={\cal T}(z)h_{\rm ini}$, which allows for simple computations of observables in terms of the primordial tensor mode amplitude.
Namely the cosmology dependence other than the initial amplitude of GWs factorizes out in the response function.

The power spectrum of biased tracers in the presence of long-wavelength GWs then has a generic form,
\begin{align}
    P_{XY}({\bf k}|h_{ij}(k_L ))
    = P_{XY}(k)\left[ 1 + \hat{k}^i\hat{k}^j h_{ij}^{\rm ini}(k_L)
    \left(R^{\rm GW}_{{\rm growth};XY}(k;k_L) + R^{\rm GW}_{{\rm dilation};XY}(k;k_L) \right)\right],
\end{align}
where $X$ and $Y$ represent tracers being considered.
For the matter auto-power spectrum ($X=Y={\rm m}$), this corresponds to the nonlinear extension of Eq.~\eqref{eq:matter_power_GWs_PT}. We will also consider the matter-halo cross-power spectrum ($X={\rm m}$ and $Y={\rm h}$) and the halo auto-power spectrum ($X=Y={\rm h}$) below.
Moreover, using these GW power spectrum responses, we can write down the $X$-$Y$-GWs (scalar-scalar-tensor) bispectrum in squeezed limit as
\begin{align}
    \lim_{k_L\to 0} B_{XY h_{(\lambda)}}(k, k', k_L) 
    = \hat{k}^i\hat{k}^j e^{(\lambda)}_{ij} \left[R^{\rm GW}_{{\rm growth};XY}(k;k_L) + R^{\rm GW}_{{\rm dilation};XY}(k;k_L) \right] P_{XY}(k) P_{h_{(\lambda)}}(k_L),
\end{align}
where we have defined the $X$-$Y$-GW bispectrum via $\langle X({\bf k})Y({\bf k}')h_{(\lambda)}({\bf k}_L)\rangle = (2\pi)^3\delta^{(3)}_{\rm D}({\bf k}+{\bf k}'+{\bf k}_L) B_{XY h_{(\lambda)}}(k,k',k_L)$ and neglected the primordial contribution.\footnote{
Although here we have neglected the primordial scalar-scalar-tensor bispectrum, its contribution to the local observables appears only at the order of ${\mathcal O}(k_L/k)^2$ and thus is subdominant for the single-field inflation (see Ref.~\cite{Pajer:2013ana}).}

Making use of the growth-dilation decomposition, we can go a little further than the perturbative result.
Because the dilation effect purely captures the coordinate transformation, using Eq.~\eqref{eq:Delta_i_beta}, we can obtain the non-perturbative result as
\begin{align}
    R^{\rm GW}_{{\rm dilation};XY}(k;k_L) =
    \beta(k_L) \frac{\partial \ln P_{XY}(k)}{\partial \ln k}.
    \label{eq:resp_dilation}
\end{align}
Eqs.~\eqref{eq:matter_power_GWs_PT} and \eqref{eq:resp_dilation} are different because former perturbative result involves the \textit{linear} power spectrum while the latter involves the \textit{nonlinear} power spectrum and can be applied to a non-linear regime.
This means that we can compute the dilation piece without running simulations even in the nonlinear regime given the slope of the nonlinear power spectrum.
On the other hand, the growth piece comes from the dynamical effect where we cannot extend the perturbative result in the nonlinear regime.
We thus need to rely on simulations in order to calibrate the growth response in the nonlinear regime.
Hence, in this paper we focus on measuring the growth term from the tidal separate universe simulations with GWs.
By comparing the measurement of the growth term of the matter auto-power spectrum in simulations with the perturbation theory prediction at quasi-nonlinear scales, we can validate our methodology using the relation:
\begin{align}
    \lim_{k\ll k_{\rm NL}}R^{\rm GW}_{\rm growth;mm}(k;k_L) =
     2\alpha(k_L).
     \label{eq:growth_pt}
\end{align}
Note that since we assume GWs are long-wavelength modes compared with scalar perturbations the perturbative result is valid only for the range of $k_L\ll k\ll k_{\rm NL}$.

\subsection{Growth response of the matter auto-power spectrum from simulations}
\label{subsec:growth_matter_GW}

In the presence of GWs of $\tau_{\rm e}$-type configuration (Eq.~\eqref{eq:h_taue}), the matter auto-power spectrum takes a form of
\begin{align}
    P_{\rm mm}({\bf k}|h_{\rm e}(k_L)) = 
    P_{\rm mm}(k)
    \left[1 + \frac23 R^{\rm GW}_{\rm growth;mm}(k;k_L)
    \left({\cal L}_2(\hat{k}_1) 
     - 
    {\cal L}_2(\hat{k}_2)\right)
     h^{\rm ini}_{\rm e} \right],
\end{align}
where $P_{\rm mm}(k)$ is the nonlinear matter power spectrum in the isotropic background,
${\cal L}_2(x)$ is Legendre polynomial of order two, $\hat{k}_1$ and $\hat{k}_2$ represent the $x$ and $y$ components of $\hat{k}$ respectively, and $h_{\rm e}^{\rm ini} \equiv- \left(h_{11}^{\rm ini} -h_{22}^{\rm ini}\right)/2 =\pm \epsilon$.
We can estimate the growth response by taking the quadrupoles of the power spectrum along both $x$ and $y$ axes:
\begin{align}
    P^{\ell_{\rm e}=2}_{\rm mm}(k|h_{\rm e}(k_L))
    & \equiv 
    P^{\ell_x=2}_{\rm mm}(k) - P^{\ell_y=2}_{\rm mm}(k)
    \\
    & =  2 P_{\rm mm}(k)  R^{\rm GW}_{\rm growth;mm}(k;k_L) h^{\rm ini}_{\rm e},
\end{align}
where $P^{\ell_i=2}_{\rm mm}(k)$ ($i=\{x,y\}$) are defined as
\begin{align}
    P^{\ell_x=2}_{XY}(k)
    & \equiv 5 \int \frac{\dd^2\hat{{\bf k}}}{4\pi} 
    P_{XY}({\bf k}) 
    {\cal L}_2(\hat{k}_1),
    \\
    P^{\ell_y=2}_{XY}(k)
    & \equiv 5 \int \frac{\dd^2\hat{{\bf k}}}{4\pi} 
    P_{XY}({\bf k}) 
    {\cal L}_2(\hat{k}_2).
\end{align}
This leads to the estimator for $R_{\rm growth;mm}(k;k_L)$ as
\begin{align}
    R^{\rm GW}_{\rm growth;mm}(k;k_L)
    =
    \frac{P^{\ell_{\rm e}=2}_{\rm mm}(k|h^{\rm ini}_{\rm e}=+\epsilon) - P^{\ell_{\rm e}=2}_{\rm mm}(k|h^{\rm ini}_{\rm e}=-\epsilon)}{4\epsilon P_{\rm mm}(k)}.
\end{align}

\begin{figure}[tb]
\centering
\includegraphics[width=\textwidth]{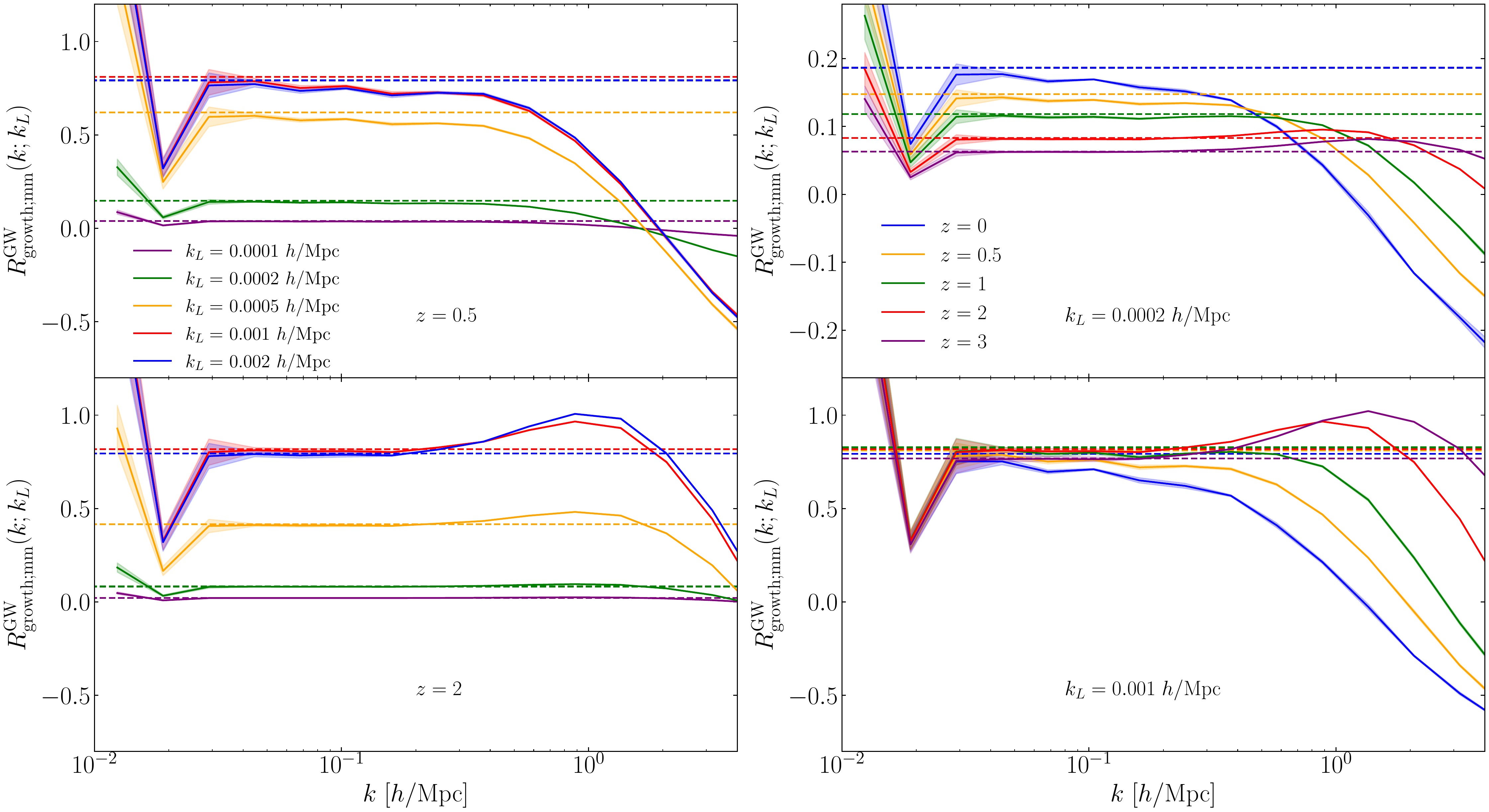}
\caption{
Growth response of matter auto-power spectrum to GWs, $R^{\rm GW}_{\rm growth}(k_L;k)$, as a function of the short-wavenmer $k$ at various wavenumvers of GWs (\textit{left}) and various redshifts (\textit{right}), measured from the simulations.
The top-left and bottom-left panels show the responses at $z=0.5$ $z=2$, respectively.
The top-right and bottom-right panels show the responses at $k_L = 0.0002~h/{\rm Mpc}$ and $k_L = 0.001~h/{\rm Mpc}$, respectively.
The dashed lines with each different color represent the predictions from the perturbation theory at each wavenumber and redshift.
}
\label{fig:Rk_gw_kLs_redshifts}
\end{figure}

\begin{figure}[t]
\centering
\includegraphics[width=\textwidth]{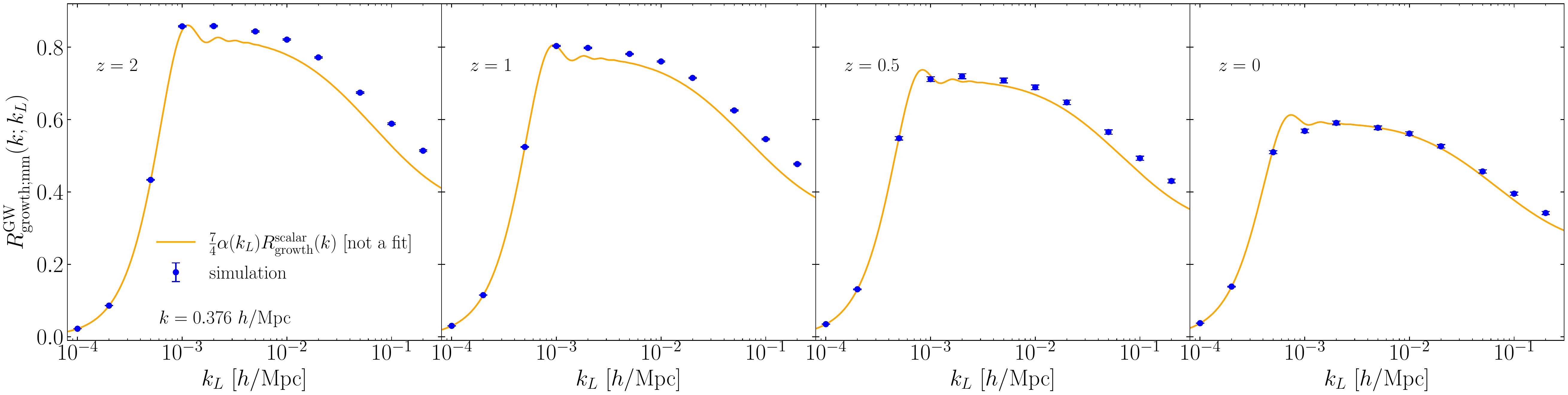}
\includegraphics[width=\textwidth]{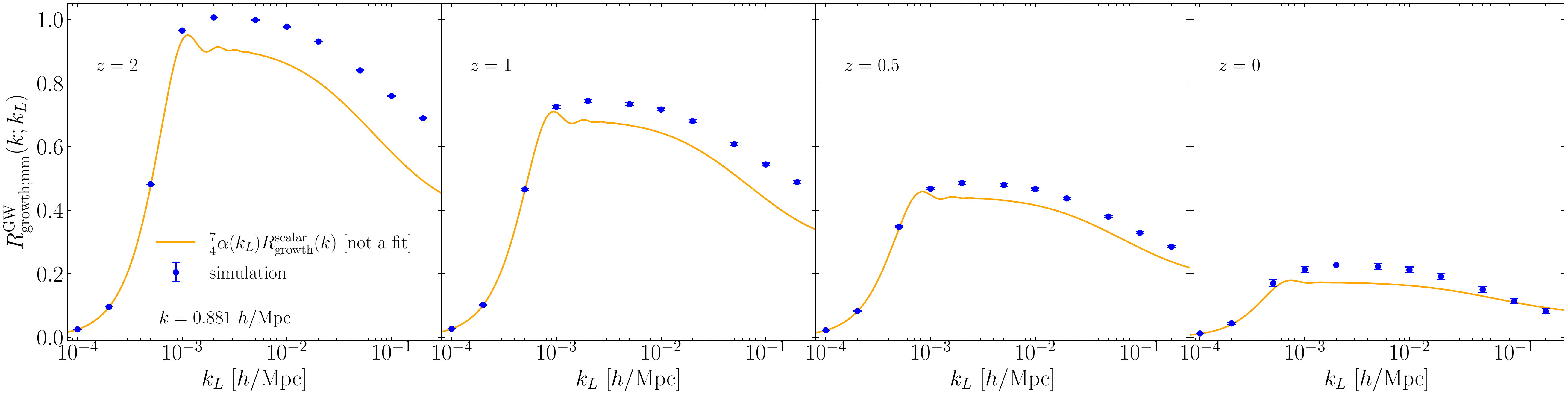}
\caption{
Growth response of matter auto-power spectrum to GWs, $R^{\rm GW}_{\rm growth}(k_L;k)$, as a function of the wavenumber of GWs $k_L$ at $z=2,1,0.5,$ and $0$ (from the left to the right), measured from the simulations.
The upper and lower rows show the results for different $k$, $k=0.376~h/{\rm Mpc}$ and $k=0.881~h/{\rm Mpc}$, respectively. 
The orange line depicts the ``rescaled growth response'' to the scalar tides, $\frac74 \alpha(k_L)R_{\rm growth}^{\rm scalar}(k)$ with $R_{\rm growth}^{\rm scalar}(k)$ measured from the tidal separate universe simulations with the scalar tides.
}
\label{fig:Rk_gw_kL}
\end{figure}

We measure the growth response of the matter auto-power spectrum to GWs, $R^{\rm GW}_{\rm growth;mm}(k;k_L)$, from the simulations which share the same initial random phase to reduce the sample variance.
Since $R^{\rm GW}_{\rm growth;mm}$ depends on the wavenumbers of the shote-mode, $k$, and of GWs, $k_L$, as well as redshift $z$, it is hard to show all the dependences in one figure. Thus, let us start by showing the measured growth response as a function of $k$ for various $k_L$ at $z=0.5$ and $z=2$ in the upper- and lower-left panels of Fig.~\ref{fig:Rk_gw_kLs_redshifts}, respectively. 
We then show it for $k_L = 0.0002h/{\rm Mpc}$ and $k_L = 0.001h/{\rm Mpc}$ at various redshifts in the upper- and lower-right panels, respectively. 
In each panel, the corresponding perturbation theory predictions, Eq.~\eqref{eq:growth_pt}, are shown in the  dashed lines.
First, onc can see the excellent agreement between the measured and predicted growth responses on large scales for all $k_L$, which verifies that our methodology to incorporate GWs in simulations works correctly.
Second, the measured response deviates from the perturbation prediction on nonlinear scales, in particular for larger $k_L$.
Although the way it deviates depends on $k_L$, the overall trend is similar among different $k_L$ values; (i) at the smallest scales ($k\gtrsim 2\ h/{\rm Mpc}$) the growth response decreases for all $k_L$ and redshifts.
(ii) At earlier redshift, the growth response is slightly enhanced compared to the perturbation theory while at lower redshift it is largely suppressed.
Third, the redshift dependence of the growth response is clearer as shown in the right panels.
The agreement between the perturbation theory and the simulation becomes worse at lower redshifts.
At $z=2$ these two are in good agreement up to $k\sim 0.2\ h/{\rm Mpc}$, while at $z=0$ the simulation results start to differ from the perturbation theory around $k\sim 0.04\ h/{\rm Mpc}$.

These tendencies we found for the tensor mode above are actually very similar to those for the case of the scalar tidal field which had been extensively studied (see App.\ref{app:scalar_tidal_bias} and Refs.~\cite{Akitsu:2020fpg,Stucker:2020fhk,Masaki:2020drx} for details).
Given the similarity of the behavior of the growth response on nonlinear scales between the scalar and tensor cases, it is natural to ask how similar are these two quantitatively.
Let us finish this subsection by answering this question.
For this purpose, we consider a ``rescaled growth response'' to the scalar tides, $\frac74\alpha(\eta;k_L)R^{\rm scalar}_{\rm growth;mm}(k)$, where the scalar tidal response $R^{\rm scalar}_{\rm growth;mm}(k)$ is introduced in Eq.~\eqref{eq:def_response_scalar} in App.~\ref{app:scalar_tidal_bias}. Since  $R^{\rm scalar}_{\rm growth;mm}(k)$ approaches $8/7$ at the large-scale limit (see Eq.~\eqref{eq:Rk_scalar_pert}), this scalar tidal response matches the tensor tidal response in the large scale limit (Eq.~\ref{eq:growth_pt}), 
\begin{align}
    \lim_{k\to 0}\frac74\alpha(\eta;k_L)R^{\rm scalar}_{\rm growth;mm}(k)=2\alpha(\eta;k_L).
\end{align}

In Fig.\ref{fig:Rk_gw_kL} 
we compare this ``rescaled growth response'' to the scalar tides with the measured tensor tidal response on nonlinear scales as a function of $k_L$ for various redshifts.
The upper and lower panels show the results for $k=0.376\ h/{\rm Mpc}$ and $k=0.881\ h/{\rm Mpc}$, respectively.
Overall, the rescaled response to the scalar tides captures the general feature of the growth response to GWs on nonlinear scales.
Such agreement can be seen for both the two wavenumbers at all the redshifts when $k_L \lesssim 10^{-3}~h/{\rm Mpc}$.
On the other hand, for $k_L \gtrsim 10^{-3}~h/{\rm Mpc}$ the response to GWs has greater values than the rescaled response, except for some $k_L$ at $z=0$.
This difference gets larger at higher redshifts and larger $k$.
These can be attributed to the different time evolution of the anisotropic scale factors in the scalar tide and GWs cases. 
The anisotropic scalar factor induced by the scalar tidal field grows monotonically in time, following the linear growth rate $D(\eta)$ at leading order, $\Delta^{\rm scalar}_i(\eta) \propto D(\eta)$ (Eq.~\eqref{eq:Delta_i_scalar}), whereas that induced by GWs has the time dependence described by $\beta(\eta;k_L)$, which is generally monotonous in time for $k_L \lesssim 10^{-3}~h/{\rm Mpc}$ while not so for $k_L \gtrsim 10^{-3}~h/{\rm Mpc}$ (see Fig.~\ref{fig:Delta_x}).
Specifically, for $k_L \gtrsim 10^{-3}~h/{\rm Mpc}$ the anisotropic scale factors from GWs reach their asymptotic value at early redshift, when the anisotropic scale factors from scalar tides are still tiny, which causes the tidal response to be stronger in these $k_L$.

\begin{figure}[t]
\centering
\includegraphics[width=\textwidth]{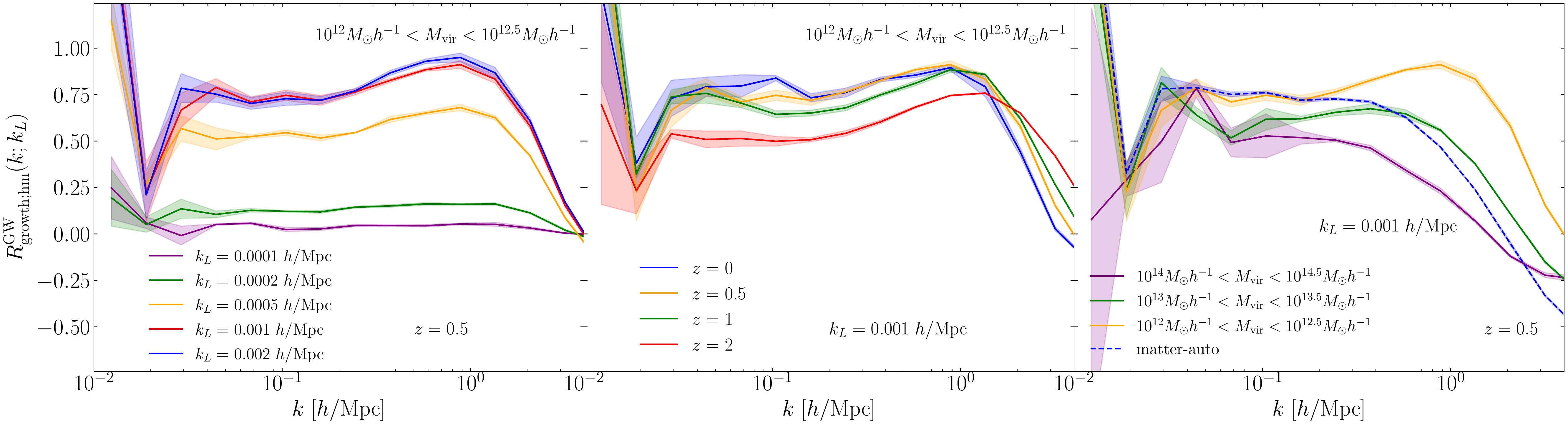}
\caption{
Growth response of the halo-matter cross-power spectrum to GWs, $R^{\rm GW;hm}_{\rm growth}(k_L;k)$, as a function of $k$ for various $k_L$ (\textit{left}), various redshifts (\textit{center}), and various halo masses (\textit{right}), measured from the simulations.
The blue dotted line in the right panel shows the growth response of the matter auto-power spectrum.
}
\label{fig:Rk_gw_hm}
\end{figure}

\begin{figure}[t]
\centering
\includegraphics[width=\textwidth]{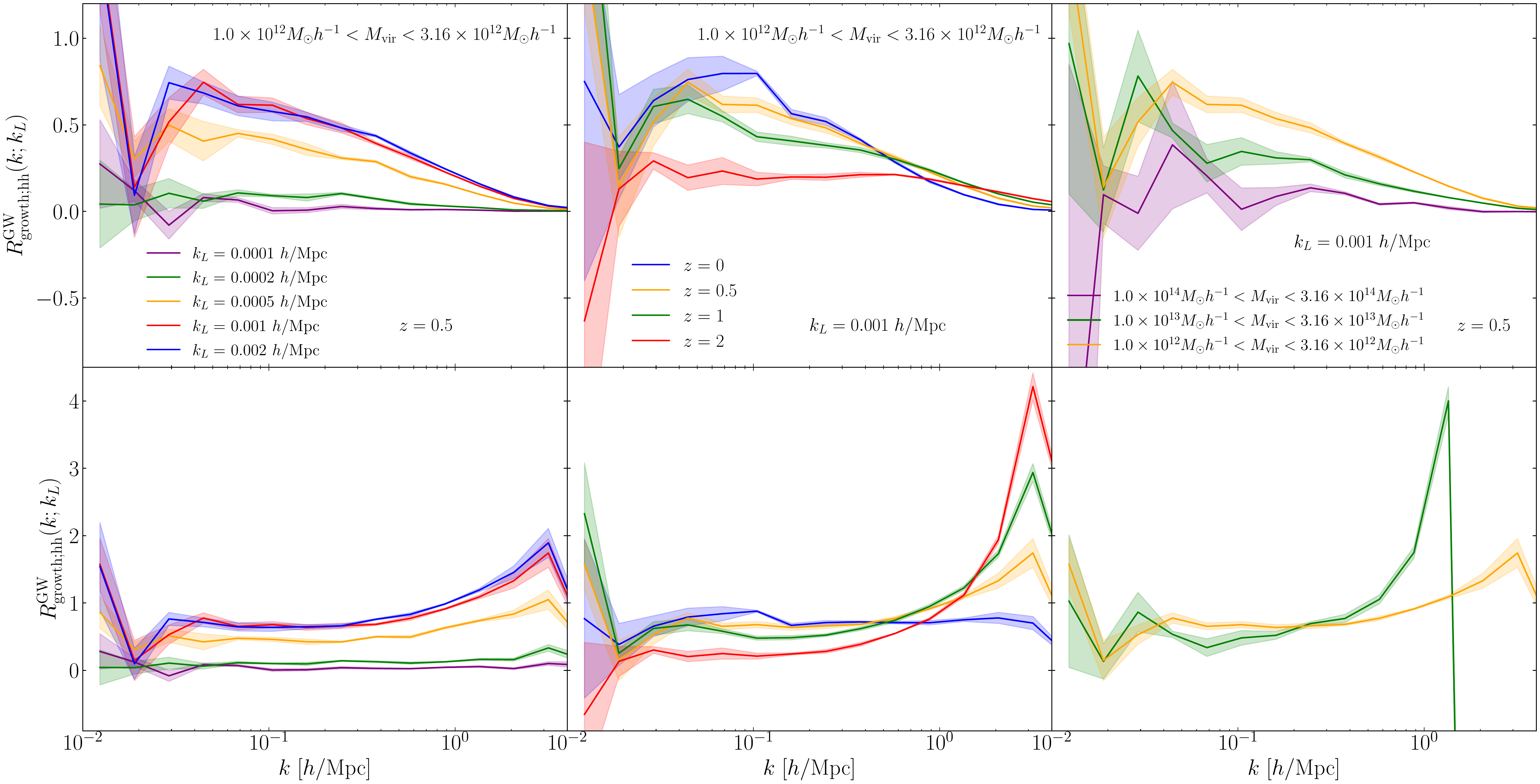}

\caption{
Similar to Fig.~\ref{fig:Rk_gw_hm} but the growth response of the halo auto-power spectrum to GWs, $R^{\rm GW;hh}_{\rm growth}(k_L;k)$ as a function of $k$ for various $k_L$ (\textit{left}), various redshifts (\textit{center}), and various halo masses (\textit{right}), measured from the simulations.
The upper and lower sets show the results normalized by the halo auto-power spectrum \textit{with} and \textit{without} a shot noise, respectively. For the latter, the shot noise contribution is subtracted assuming the Poisson distribution. 
In the lower-right panel, we do not show the result for the mass bin, $10^{14} M_\odot/h < M_{\rm vir} < 10^{14.5} M_\odot/h$ because it is too noisy. 
}
\label{fig:Rk_gw_hh}
\end{figure}

\subsection{Growth responses of the halo-matter cross- and halo auto-power spectra from simulations}
\label{subsec:growth_halo_GW}

The halo-matter cross- and halo auto-power spectra in a local region are also affected by GWs.
The growth responses of these power spectra to GWs can be estimated from the simulations in the same way as the matter auto-power spectrum,
\begin{align}
    R^{\rm GW}_{\rm growth;XY}(k;k_L)
    =
    \frac{P^{\ell_{\rm e}=2}_{\rm XY}(k|h^{\rm ini}_{\rm e}=+\epsilon) - P^{\ell_{\rm e}=2}_{\rm XY}(k|h^{\rm ini}_{\rm e}=-\epsilon)}{4\epsilon P_{\rm XY}(k)}, 
    \label{eq:R_GW_halo}
\end{align}
with $XY \in \{{\rm hm}, {\rm hh}\}$.

In Fig.~\ref{fig:Rk_gw_hm} we show the growth response of the halo-matter cross-power spectrum to GWs as a function of the wavenumber of the short modes.
The left and middle panels focus on the fixed halo mass ($10^{12} M_\odot/h < M_{\rm vir} < 10^{12.5} M_\odot/h$), and compares the growth responses for various wavenumbers of GWs at $z=0.5$ in the left panel and various redshfits at $k_L = 0.001~h/{\rm Mpc}$ in the middle panel.
As for the $k_L$-dependence, the response tends to be greater for larger $k_L$, similar to the case of the matter auto response.
On the other hand, the redshift dependence is slightly different from the matter auto case.
For the halo-matter cross-power spectrum, the growth response at late redshifts ($z=0$ and $z=0.5$) gets slightly enhanced compared to the linear regime and persists up to $k \lesssim 1~h/{\rm Mpc}$, whereas the matter auto response starts to decrease around $k \simeq 0.1~h/{\rm Mpc}$.
This can be clearly seen in the right panel of Fig.~\ref{fig:Rk_gw_hm}, which directly compares the growth response of the matter auto power-spectrum to that of the halo-matter cross-power spectrum with various halo masses at $z=0.5$ and $k_L = 0.001~h/{\rm Mpc}$.
First, it turns out that the scale-dependence ($k$-dependence) of the growth response varies with the halo mass.
The enhancement of the growth response around $k \simeq 1~h/{\rm Mpc}$ is greater for less massive halos.
This is mainly due to the normalization.
Here we define the response with respect to the halo-matter cross-power spectrum in the fiducial simulation.
However if we define the response with respect to the matter auto-power spectrum, the response of more massive halos gets more amplified in the nonlinear regime.
Second, the difference of the responses between the matter auto and halo-matter cross cases stems from the halo biases.
In particular at largest scales it should be explained by the halo tidal bias induced by GWs, which we investigate in the next section.

Next, let us focus on the growth response of the halo auto-power spectrum.
To obtain it, for the denominator of Eq.~(\ref{eq:R_GW_halo}) we use measurements of the halo auto-power spectrum with and without the shot noise contribution. For the latter, the shot noise contribution is subtracted assuming the Poisson distribution, $1/\bar{n}_{\rm h}$.
The upper and lower rows of Fig.~\ref{fig:Rk_gw_hh}
show the resulting response of the halo auto-power spectrum as a function of $k$ with and without the shot noise, respectively. 
As with Fig.~\ref{fig:Rk_gw_hm}, 
the left and middle columns show the responses with the fixed mass range $10^{12} M_\odot/h < M_{\rm vir} < 10^{12.5} M_\odot/h$ for various wavenumber of GWs at $z=0.5$ and for $k_L = 0.001~h/{\rm Mpc}$ at various redshfits, respectively, while the right columns show the result with  various halo masses for $k_L = 0.001~h/{\rm Mpc}$ at $z=0.5$.
Overall, the $k_L$, redshift, and halo mass dependencies of the responses for the halo auto spectra both with and without the shot noise are similar to that for the halo-matter cross spectrum.
In the upper set of Fig.~\ref{fig:Rk_gw_hh}, the responses approach zero on small scales where the halo auto-power spectra are dominated by shot noises.
On the contrary, the peaky features around $k\sim 2~h/{\rm Mpc}$ in the lower set reflects the non-Poissonian behavior of the halo shot noise 
due to, e.g., the exclusion effect~\cite{Smith:2006ne,Hamaus:2010im,Baldauf:2013hka}.
Therefore, the simple Poissonian shot-noise removal leads to zero-crossing or negative, unphysical halo auto-power spectrum on small scales, making the responses peaky.
Nonetheless, up to $k\sim 0.5~h/{\rm Mpc}$, where the shot noise contribution is still small, the responses are not suppressed unlike the matter auto case.
These trends seen in the response of the halo-matter cross- and halo auto-power spectra are the same as those in the scalar tidal field case (see App.~\ref{app:scalar_tidal_bias}).

\section{Halo tidal bias induced by GWs}
\label{sec:tidal_bias}

\begin{figure}[tb]
\centering
\includegraphics[width=0.9\textwidth]{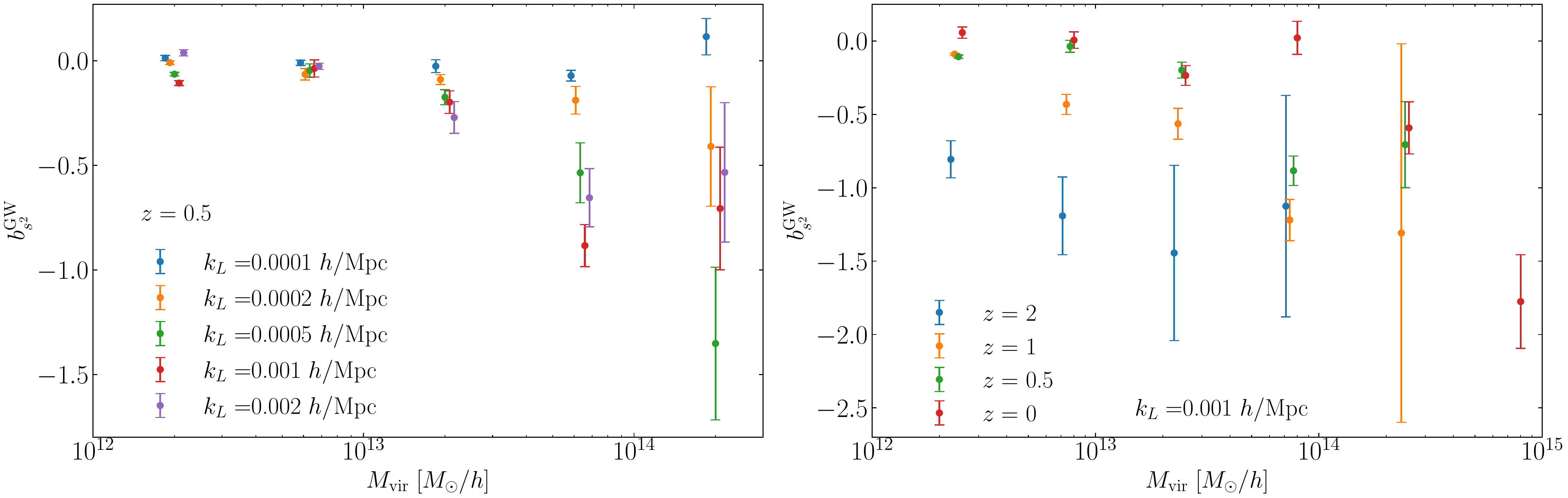}
\caption{
Halo tidal bias induced by GWs as a function of halo mass for various wavenumber of GWs at $z=0.5$ (the left panel) and at various redsfhits for $k_L = 0.001~h/{\rm Mpc}$ (the right panel).
}
\label{fig:bs2_GW_mass}
\end{figure}

As discussed in the previous section, the long-wavelength GWs can affect the halo density field.
In the perturbative regime, this effect should be characterized in terms of the halo bias. 
When only scalar perturbations are considered, which is the standard setup, the halo density field can be expanded up to the second order as
\begin{align}
    \delta_{\rm h} = b_1 \delta + \frac12 b_2 \delta^2 + b_{s^2} s^2,
    \label{eq:bias_expansion}
\end{align}
where $s^2 = s_{ij}s^{ij}$ with $s_{ij}\equiv (\partial_i\partial_j/\partial^2 - \deltaK_{ij}/3)\delta$.
Here $b_{s^2}$ is called the tidal bias that captures the effect of the tidal fields on the halo density field.
Since long-wavelength GWs are locally indistinguishable from tidal fields induced by the scalar perturbations, it is natural to expect that there is a tidal bias induced by GWs as well.
Then, at the linear order of GWs, the halo density field should acquire the following contribution from GWs,
\begin{align}
    \delta_{\rm h}^{(2)}(\eta) \supset b^{\rm GW}_{s^2}(\eta;k_L) s^{ij}(\eta) h_{ij}^{\rm ini}(k_L),
\end{align}
where $b_{s^2}^{\rm GW}$ is the tidal bias coefficient induced by the coupling between tidal fields induced by GWs and scalar density perturbations.
Note that we define $b_{s^2}^{\rm GW}$ with respect to the GWs at the \emph{initial} epoch instead of the \emph{same} time, unlike the biases defined with respect to the scalar perturbations.
Because the GWs are additional degrees of freedom to the scalar adiabatic perturbations, 
we expect the tidal bias to depend on the wavenumber of GWs.
Collecting the second order pieces in the matter density field and halo density field,
we find the tree-level halo-matter-GWs bispectrum in the squeezed limit as 
\begin{align}
    \lim_{k_L\to 0}B_{{\rm hm}h_{(\lambda)}}(k,k',k_L) 
    = \hat{k}^i\hat{k}^j e^{(\lambda)}_{ij} \left[ 2b_1\alpha(\eta;k_L) + 2b_{s^2}^{\rm GW}(\eta;k_L) \right] P_{\rm lin}(k) P_{h_{(\lambda)}}(k_L),
\end{align}
where we have omitted the dilation piece.
This implies that the local halo-matter power spectrum in the presence of long-wavelength GWs is
\begin{align}
    P_{\rm hm}({\bf k}|h_{ij}(k_L)) 
    =\left[ b_1 
    + \left[2b_1\alpha(\eta; k_L)+2b^{\rm GW}_{s^2}(\eta;k_L)\right]k^ik^j h_{ij}^{\rm ini}\right]P_{\rm lin}(k).
    \label{eq:Phm_GW}
\end{align}
Thus, we can estimate $b_{s^2}^{\rm GW}$ from the growth response of the halo-matter power spectrum to GWs, which is presented in the previous section.
Although there are several ways to estimate $b_{s^2}^{\rm GW}$ from the growth response of the halo-matter power spectrum,
we use the following way to reduce the uncertainty of $b_1$ and  the sample variance.
We first define the local linear bias estimator as a ratio of the halo-matter and matter auto-power spectra including the quadrupoles measured in the simulations with GWs,
\begin{align}
    \hat{b}^{\ell_{x,y}}_1(k;\pm \epsilon) \equiv \frac{P^{\ell = 0}_{\rm hm}(k;\pm\epsilon) + P^{\ell_{x,y} = 2}_{\rm hm}(k;\pm\epsilon)}{P^{\ell = 0}_{\rm mm}(k;\pm\epsilon) + P^{\ell_{x,y} = 2}_{\rm mm}(k;\pm\epsilon)}
    = \frac{b_1 \pm (2b_1\alpha(\eta;k_L) + 2b_{s^2}^{\rm GW})\epsilon}{1\pm2\alpha(\eta;k_L)\epsilon}
    \simeq \tilde{b}_1(k) \pm  2 \tilde{b}_{s^2}^{\rm GW}(k)\epsilon,
    \label{eq:local_b1_GW}
\end{align}
Therefore using this local linear bias, the estimator for $b_{s^2}^{\rm GW}$ is now 
\begin{align}
    \tilde{b}^{\rm GW}_{s^2}(k) = \frac12\left[\frac{\hat{b}_1^{\ell_{x}}(k;+\epsilon) - \hat{b}_1^{\ell_{x}}(k;-\epsilon)}{4\epsilon}
    + \frac{\hat{b}_1^{\ell_{y}}(k;+\epsilon) - \hat{b}_1^{\ell_{y}}(k;-\epsilon)}{4\epsilon}\right].
    \label{eq:bs2_GW_estimator}
\end{align}
We estimate $b_{s^2}^{\rm GW}$ utilizing the $\chi^2$ statistic, defined as $\chi^2 = \sum_{k=k_{\rm min}}^{k_{\rm max}}[b_{s^2}^{\rm GW} - \tilde{b}_{s^2}^{\rm GW}(k)]^2/\sigma^2_{b_{s^2}^{\rm GW}}(k)$, where $\sigma^2_{b_{s^2}^{\rm GW}}(k)$ is the variance of $\tilde{b}_{s^2}^{\rm GW}(k)$ at each $k$-bin measured from simulations.
Adopting $k_{\rm max} = 0.08 ~h/{\rm Mpc}$, we obtain the best-fitting value of $b_{s^2}^{\rm GW}$ and its uncertainty by minimizing $\chi^2$.
We restrict the measurement of $b_{s^2}^{\rm GW}$ to a range of $k_L \leq 0.002~h/{\rm Mpc}$ because we rely on the perturbative results, which are valid only when $k_L \ll k$.

\begin{figure}[t]
\centering
\includegraphics[width=\textwidth]{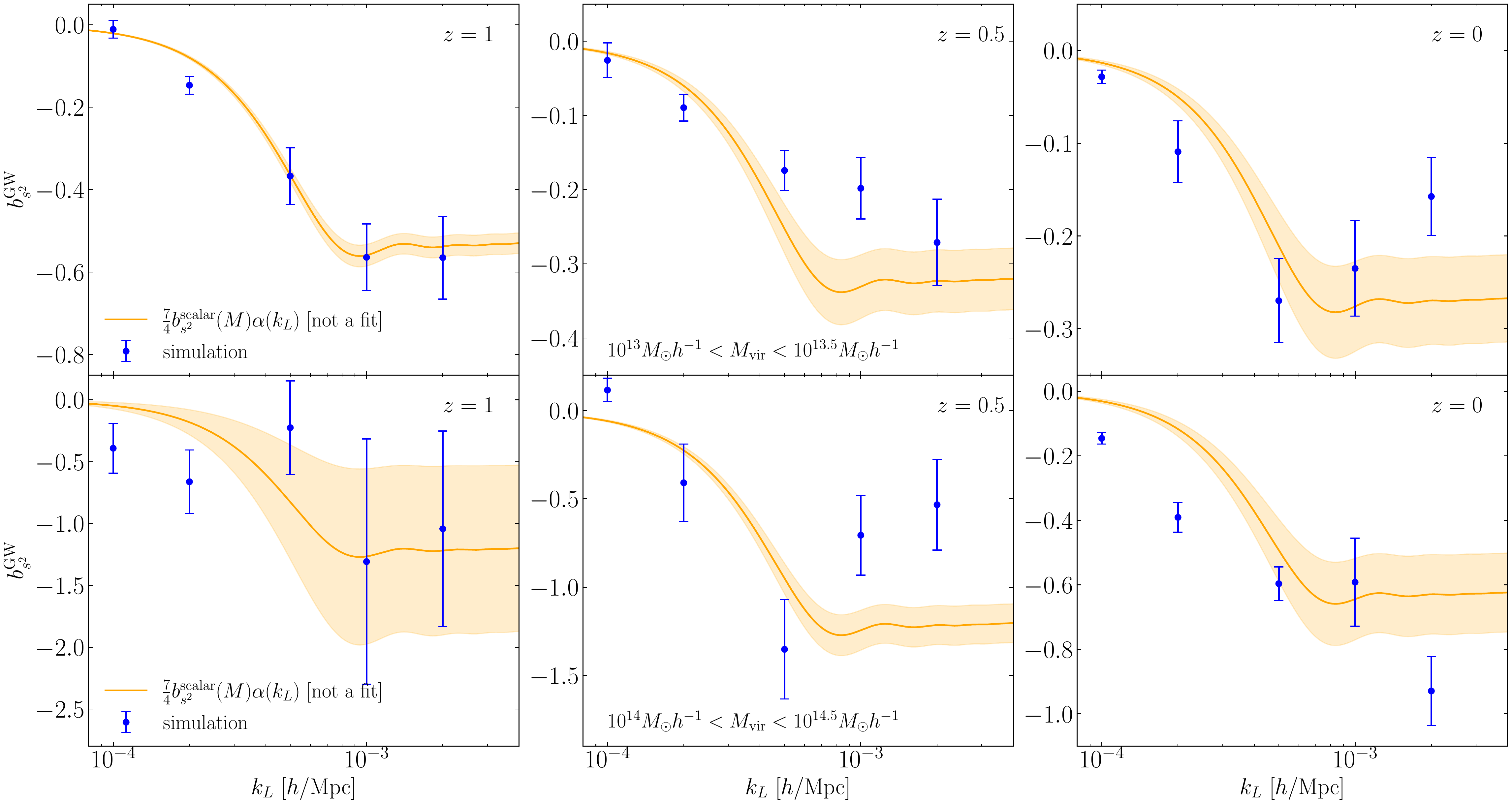}
\caption{
Upper panels: Halo tidal bias induced by GWs as a function of wavenumber of GWs with halo mass of $10^{13} ~M_\odot/h < M_{\rm vir} < 10^{13.5} ~M_\odot/h$. 
Lower panels: Same as upper panels but with halo mass of $10^{14} ~M_\odot/h < M_{\rm vir} < 10^{14.5} ~M_\odot/h$.
From the left to right , we show the results at $z=1$, $0.5$ and $0$. 
The blue points with error bars are the measurement from the simulations.
The orange line represents the ansatz $b_{s^2}^{\rm GW}(k_L; M) = \frac74 b_{s^2}^{\rm scalar}(M) \alpha(k_L)$ with the shaded region being the error coming from $b_{s^2}^{\rm scalar}(M)$.
}
\label{fig:bs2_GW_kL}
\end{figure}

Fig.~\ref{fig:bs2_GW_mass} shows the tidal bias from GWs determined in this way as a function of halo mass.
In the left panel we plot the result for various wavenumbers of GWs at $z=0.5$, and in the right panel we plot the result for various redshifts at $k_L = 0.001~h/{\rm Mpc}$.
Regardless of the wavenumber of GWs and redshift, more massive halos have greater absolute values of $b_{s^2}^{\rm GW}$, which has the same trend as the usual tidal bias induced by the scalar tidal fields.
In addition to the mass dependence, the redshift dependence is also similar to the scalar tidal bias case;
the absolute value of the tidal bias is larger at higher redshift.
On the other hand, in contrast to the case of the scalar tidal bias, the tidal bias from GWs has a particular wavenumber dependence as shown in the left panel of Fig.~\ref{fig:bs2_GW_mass}.
This wavenumber-dependence (or scale-dependence) inherits from the wavenumber-dependent transfer function of GWs while the growth function for the scalar density fluctuations is independent of the wavenumber (Eq.~\eqref{eq:growth_func}).
To look at this scale dependence in more detail, Fig.~\ref{fig:bs2_GW_kL} displays $b_{s^2}^{\rm GW}$ as a function of $k_L$ for several redshifts and halo masses.
We compare the measurements with the following ansatz:
\begin{align}
    b_{s^2}^{\rm GW}(\eta;k_L) = \frac74  \alpha(\eta;k_L) b^{\rm scalar}_{s^2}(\eta),
\end{align}
where $b_{s^2}^{\rm scalar}$ is the tidal bias induced by the scalar tides, introduced in Eq.~\eqref{eq:bias_expansion} (see also App.~\ref{app:scalar_tidal_bias}).
This ansatz is motivated by the fact that (i) the tidal bias term stems from the coupling of tidal fields, which is given by $s^{ij}(z)s_{ij}(z)$ for the scalar tide case ($b_{s^2}^{\rm scalar}$) while $s^{ij}(z)h^{\rm ini}_{ij}$ for the tensor tide case ($b_{s^2}^{\rm GW}$),
and (ii) the second-order matter density induced by tidal fields is $\frac47 s^{ij}(z)s_{ij}(z)$ for the scalar tides case while $\alpha(z;k_L) s^{ij}(z) h^{\rm ini}_{ij}(k_L)$ for the tensor tides case.
Despite the large error bars, overall the measurements are well explained by this ansatz.
This result suggests that we do not have to introduce a new free bias parameter for GWs at leading order once $b_{s^2}^{\rm scalar}$ is known.

\section{Intrinsic alignments induced by GWs}
\label{sec:IA_GW}

\begin{figure}[tb]
\centering
\includegraphics[width=\textwidth]{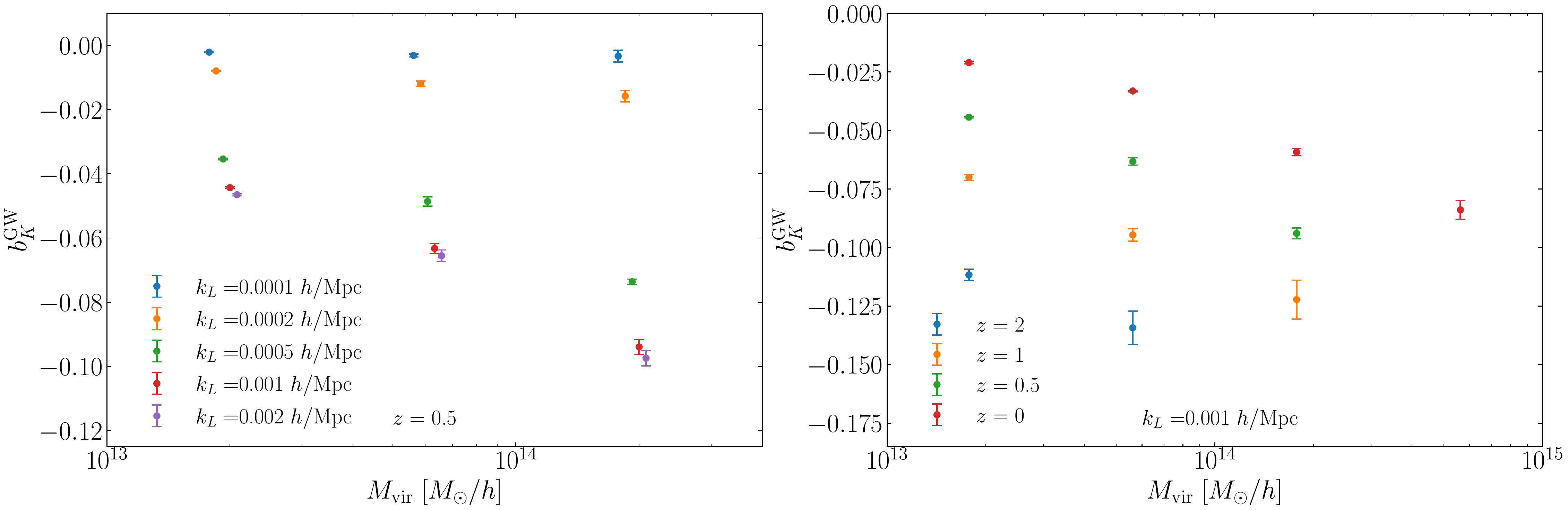}
\caption{
Linear shape bias induced by GWs as a function of halo mass.
The left and right panels show $b_K^{\rm GW}$ for various wavenumber of GWs at $z=0.5$ and for various redshifts at $k_L = 0.001~h/{\rm Mpc}$, respectively.
}
\label{fig:bK_GW_mass}
\end{figure}

As we discussed so far, the tidal fields contribute to the density field only at the second order because tidal fields are a tensor while density fields are a scalar.
Conversely, the tidal fields should contribute to tensor quantities at linear order.
One well-known observable of tensor quantities in large-scale structure is the intrinsic alignments of halo (or galaxy) shapes~\cite{Catelan:2000vm,Hirata:2004gc}.
The deviation of the halo intrinsic shape from sphere is characterized by ellipticity at the lowest order.
In other words, the halo shape can be described by the trace-free rank-2 tensor in three-dimensional space, $\gamma_{ij}$.

The linear alignment model of the intrinsic alignment relates the halo shape with the tidal field as follows:
\begin{align}
    \gamma_{ij}(\eta) &= b^{\rm scalar}_K(\eta) s_{ij}(\eta),
    \label{eq:IA_scalar}
\end{align}
where $b^{\rm scalar}_K$ is the linear alignment coefficient or the linear shape bias, which captures the sensitivity or the response of halo or galaxy shape to the tidal field.
Because long-wavelength GWs are locally equivalent to the tidal field, we naturally expect that the halo shapes are also aligned by GWs, namely
\begin{align}
    \gamma_{ij}(\eta;k_L) = b_K^{\rm GW}(\eta;k_L) h^{\rm ini}_{ij}(k_L),
    \label{eq:IA_GW}
\end{align}
where we have introduced the linear shape bias $b_K^{\rm GW}$ for GWs, in analogy with the tidal bias, $b^{\rm scalar}_K$.
While $b^{\rm scalar}_K$ is defined as a response of the halo shape with respect to the scalar tidal field at the \textit{same} time, $b^{\rm scalar}_K$ is defined as that with respect to GWs at the \emph{initial} time when the GWs are frozen, as in the tidal bias case.

\begin{figure}[tb]
\centering
\includegraphics[width=\textwidth]{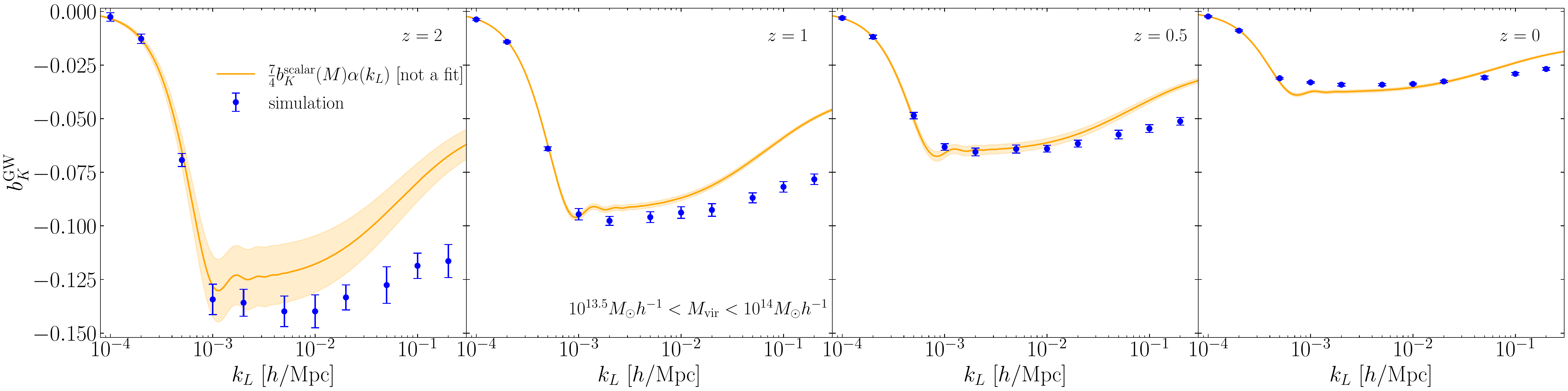}
\caption{
Linear shape bias induced by GWs as a function of the wavenumber of GWs for the halos with $10^{13.5} ~M_\odot/h < M_{\rm vir} < 10^{14} ~M_\odot/h$.
From the left to the right panels we show the results at $z=2,1,0.5,$ and $0$, respectively.
The orange line depicts the ansatz introduced in Eq.~\eqref{eq:bK_ansatz}, $b_K^{\rm GW}(k_L;M,z) = \frac74\alpha(k_L,z) b_K^{\rm scalar}(M,z)$, and is not a fitting.
The orange dashed region corresponds to the 1-$\sigma$ error of $b_K^{\rm scalar}$.
}
\label{fig:bK_GW_kL}
\end{figure}

The linear shape bias introduced above can be efficiently measured in the tidal separate universe simulation~\cite{Akitsu:2020fpg,Stucker:2020fhk}, as the linear halo bias can be obtained precisely in the isotropic separate universe simulation~\cite{Li:2015jsz,Baldauf:2015vio,Lazeyras:2015lgp}.
In this paper, we define the halo shape by its reduced inertia tensor,
\begin{align}
    J_{ij} = \sum_{n=1}^{N_{\rm p}} \frac{x_{n,i}x_{n,j}}{x_n^2},
\end{align}
where $N_{\rm p}$ is the number of particles in the halo and $x_{n,i}$ is the $i$-th component of the particle location  with respect to the halo center.
Note that here $x$ represents the distance measured in the \textit{isotropic} background, i.e., the physical coordinates.
As $\gamma_{ij}$ is regarded as the trace-free part of $J_{ij}$,
the linear alignment model states that this halo shape can be written as
\begin{align}
    J_{ij} = J_0\left[ \frac13 \deltaK_{ij} + \gamma_{ij} \right] = J_0\left[ \frac13 \deltaK_{ij} + b_K K_{ij} \right],
\end{align}
where $J_0$ is the normalization, for which we use the trace part of $J_{ij}$: $J_0 = {\rm Tr}[J_{ij}]$, and $K_{ij}$ is the tidal field of either $s_{ij}$ or $h_{ij}^{\rm ini}$.
Therefore, we can measure $b_K$ as a response of the one-point function of the ellipticity to the tidal field.
Specifically, introducing the following quantity
\begin{align}
    J_{\rm e} \equiv \frac{J_{11}-J_{22}}{2}, 
\end{align}
$b_K$ can be obtained as 
\begin{align}
    b_K(M,z) = \frac{J_{\rm e}(M,z|K_{\rm e}^{\rm ini}=+\epsilon) - J_{\rm e}(M,z|K_{\rm e}^{\rm ini}=-\epsilon) }{2\epsilon J_0(M,z)},
    \label{eq:bK_estimator}
\end{align}
where $K_{\rm e}^{\rm ini} = h_{\rm e}^{\rm ini} = \left(h_{11}^{\rm ini} -h_{22}^{\rm ini}\right)/2$ for the tensor tides and $K_{\rm e}^{\rm ini} = \left(s_{11}(z) -s_{22}(z)\right)/2$ for the scalar tides.

Fig.~\ref{fig:bK_GW_mass} shows $b_K^{\rm GW}$ as a function of halo mass for various wavenumbers of GWs at $z=0.5$ (the left panel) and for $k_L = 0.001~h/{\rm Mpc}$ at various redshifts (the right panel).
First, we find that GWs influence the halo shape, implying that indeed intrinsic alignments can be induced by GWs.
Compared with the measurement of $b_{s^2}^{\rm GW}$, the measurement of $b_K^{\rm GW}$ has much greater signal-to-noise ratio as we use the one-point function to measure $b_K^{\rm GW}$ while $b_{s^2}^{\rm GW}$ is measured from the large-scale limit of the power spectrum responses.
Second, the halo-mass and redshift dependences of $b_K^{\rm GW}$ are similar to those of $b_K^{\rm scalar}$; that is, massive halos tend to more strongly align with GWs and the strength of the alignment at fixed halo mass decreases as the redshift gets smaller (see Fig.~\ref{fig:bK_scalar} and Ref.~\cite{Akitsu:2020fpg} for details).\footnote{
Note, however, that for the tensor case $b_K^{\rm GW}(z)$ is defined with respect to the \emph{initial} amplitude of GWs and it thus represents the strength of the alignment with respect to $h_{ij}^{\rm ini}$, while for the scalar tides case, $b_K^{\rm scalar}(z)$ represents the strength of the alignment with respect to $s_{ij}(z)$.
Taking this difference into account, however, does not change the trend of the redshift dependence though,
since $s_{ij}\propto D$ that increases monotonically with time.
}
Third, the unique feature in the case of GWs is that $b_K^{\rm GW}$ is wavenumber-dependent (or scale-dependent), as in the cases with the power spectrum response and the halo tidal bias.
Fig.~\ref{fig:bK_GW_kL} plots $b_K^{\rm GW}$ as a function of the wavenumber of GWs for the halos with $10^{13.5} ~M_\odot/h < M_{\rm vir} < 10^{14} ~M_\odot/h$ at $z=2,1,0.5, 0$ to directly demonstrate this scale dependence.
The scale dependence is clearly seen because of the high $S/N$.

\begin{figure}[tb]
\centering
\includegraphics[width=\textwidth]{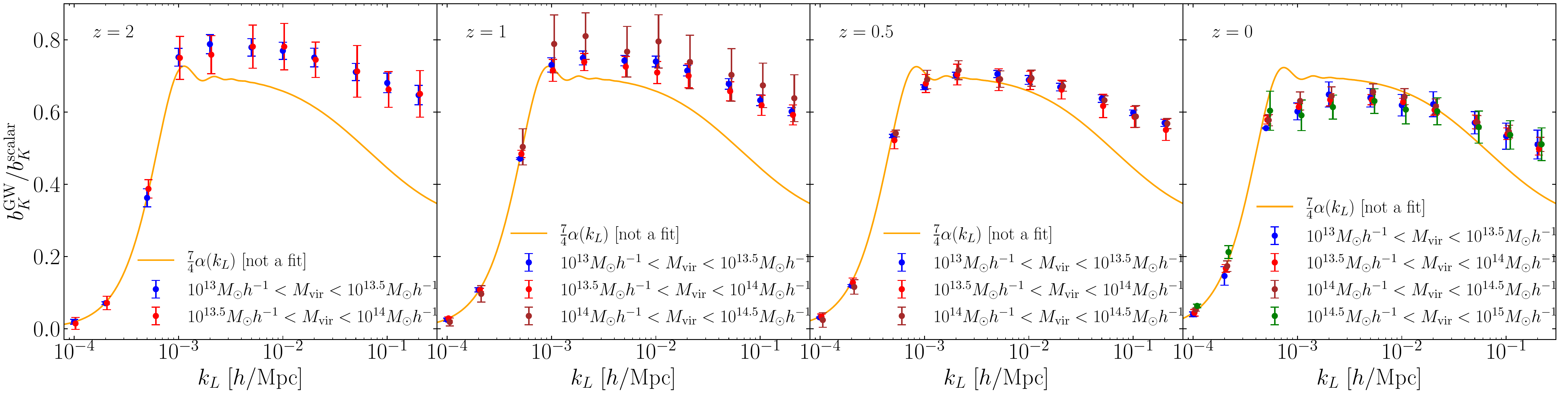}
\caption{
Ratio of the linear shape bias induced by GWs to the one induced by scalar tidal fields as a function of the wavenumber of GWs for various halo masses.
From the left to the right, we show the results at $z=2,1,0.5,0$, respectively.
The orange line depicts the ansatz introduced in Eq.~\eqref{eq:bK_ansatz}, $b_K^{\rm GW}(k_L;M,z)/b_K^{\rm scalar}(M,z) = \frac74\alpha(k_L,z) $, and is not a fitting.
}
\label{fig:bK_GW_mass_kL}
\end{figure}

In Fig.~\ref{fig:bK_GW_kL} we compare the measurements with the following ansatz for the intrinsic alignment from GWs~\cite{Schmidt:2013gwa},\footnote{
The original ansatz considered in Ref.~\cite{Schmidt:2013gwa} is $b_K^{\rm GW}(\eta;k_L)  = \frac72 \alpha(\eta;k_L) b_K^{\rm scalar}(\eta)$ (see their Eq.~(74)). However, the prefactor should be $7/4$ given the factor of two that comes from exchanging long-mode and short-mode in the second-order coupling ($\delta^{(2)} = \frac27 s_{ij}s_{ij} \to \delta^{(2)} = \frac47 s^{\rm long}_{ij}s^{\rm short}_{ij}$).}
\begin{align}
    b_K^{\rm GW}(\eta;k_L) = \frac74 \alpha(\eta;k_L) b_K^{\rm scalar}(\eta).
    \label{eq:bK_ansatz}
\end{align}
It is natural to assume that the halo shape is determined by the local tidal environment around the halo, which is affected by long-wavelength tidal field via nonlinear mode-coupling.
The influence of the long-wavelength tides on the small-scalar tides is captured by the response function that is discussed in Sec.~\ref{sec:power_response}.
In particular, in Fig.~\ref{fig:Rk_gw_kL}
we examine the relation between the response function to the scalar tides and GWs and find that $R_{\rm growth;mm}^{\rm GW}(k;k_L) = \frac74 \alpha(\eta;k_L)R_{\rm growth;mm}^{\rm scalar}(k)$ is a good approximation even in the nonlinear regime.
As these responses serve as the amplitude of the small-scale tides induced by the large-scale tides,
we expect that $\gamma^{\rm scalar}_{ij} \propto R_{\rm growth;mm}^{\rm scalar} s_{ij}$ and $\gamma^{\rm GW}_{ij} \propto R_{\rm growth;mm}^{\rm GW} h^{\rm ini}_{ij}$, leading to the above ansatz.
The ansatz of Eq.~\eqref{eq:bK_ansatz} is shown by the orange lines in Fig.~\ref{fig:bK_GW_kL}.
Remarkably, the measurements are well explained by this prediction for all the redshifts.
Fig.~\ref{fig:bK_GW_mass_kL} investigates if this agreement holds for all halo mass by displaying the ratio of the linear shape biases, $b_K^{\rm GW}/b_K^{\rm scalar}$, which is equal to $\frac74 \alpha$ in the ansatz regardless of halo mass.
It turns out that the trend seen in Fig.~\ref{fig:bK_GW_kL} remains the same for all halo masses.
One important point that follows from this agreement is that the process to determine the halo shape is not local in time.
For, if the halo shape responds to the tidal field locally in time, meaning that the halo shape is related to the instantaneous tides: $\gamma^{\rm scalar}_{ij}(\eta) \propto s_{ij}(\eta)$ and $\gamma^{\rm GW}_{ij}(\eta) \propto \tau_{ij}(\eta)$, we expect $b_K^{\rm GW}(\eta) = b_K^{\rm scalar}(\eta) T(\eta;k_L)/a(\eta)$ with $T$ defined in Eq.~\eqref{eq:CFC_tides}; however this is not the case.

Let us finish this section by considering possible reasons that cause the difference between the measurements and the ansatz. 
First, the deviation is also observed around $k_L \sim 10^{-3}~h/{\rm Mpc}$ at $z=0$.
This might be inherited from the large difference in the response functions to the scalar and tensor tides shown in the upper row of Fig.~\ref{fig:Rk_gw_kL},
which implies that relatively large-scale tides are more responsible for halo shapes than halo-scale tides.
Second, the deviation of the measurements from the ansatz gets larger as $k_L$ increases.
This trend is also consistent with the responses of the matter power spectrum in Fig.~\ref{fig:Rk_gw_kL}.
At the same time, this could be partly because the approximation we employ in this paper is no longer valid for these large $k_L$.
In other words, GWs with larger-$k_L$ cannot be seen as a uniform tidal field even in the halo formation region, given that the Lagrangian halo radius $R_M = \left(4\pi\bar{\rho}_{\rm m}/3M_{\rm vir}\right)^{1/3}$ is close to the wavelength of GWs.
Specifically, ignoring the curvature of GWs can leads the corrections to $b_K^{\rm GW}$, which scales as ${\cal O}(k_L^2R_M^2)$. In the case of $M_{\rm vir} = 1.0\times 10^{13} M_\odot/h$ and $k_L = 0.1~ h/{\rm Mpc}$,
for example, the correction is about $(0.1\cdot 3)^2\sim 0.1$, which cannot be negligible.
In order to accurately study the impact of such GWs, a different technique is necessary.

We also note that the results for high-$k_L$ GWs can be verified by analyzing the standard $N$-body simulation with staring an anisotropic initial power spectrum because the effect of high-$k_L$ GWs is almost encoded in the initial conditions.
As shown in Fig.~\ref{fig:Delta_x}, the anisotropic scale factors induced by high-$k_L$ GWs ($k_L \gtrsim 10^{-2}~h/{\rm Mpc}$) reach their asymptote already at $z_{\rm ini}$ and thus their effect in the late time can be absorbed into the overall time-independent rescaling of scale factors, implying that it does not have a physical effect on structure formation.
In other words, for high-$k_L$ GWs halos are formed in absence of a large-scale tidal field but with anisotropic small-scale modes, while for low-$k_L$ GWs halos are formed in an evolving tidal field, in analogy to the difference in $b_1$ (the response to the large-scale density field) and $b_\phi$ (the response to the change of $\sigma_8$) in the density case~\cite{Baldauf:2011bh,Desjacques:2016bnm}.
This study is beyond our paper and we leave it for future work.

\section{Discussion}
\label{sec:Discussion}

In this paper, we have quantified the impact of GWs on large-scale structure by means of the tidal separate universe simulation.
To the best of our knowledge, this is the first study for the effect of GWs on nonlinear structure formation using $N$-body simulations.
We found that GWs indeed influence nonlinear structure formation in both the clustering statistics and the intrinsic alignment of halo shapes.
Our main finding is that the impact of GWs on large-scale structure can be well described by combining the impact of the scalar tides with the perturbation theory.
Specifically, the halo tidal bias and linear shape bias induced by GWs, $b_{s^2}^{\rm GW}$ and $b_{K}^{\rm GW}$ respectively, can be approximated by $b_X^{\rm GW}(k_L) = \frac74 \alpha(k_L) b_X^{\rm scalar}$ where $X = \{s^2,K\}$.

\begin{figure}[tb]
\centering
\includegraphics[width=0.6\textwidth]{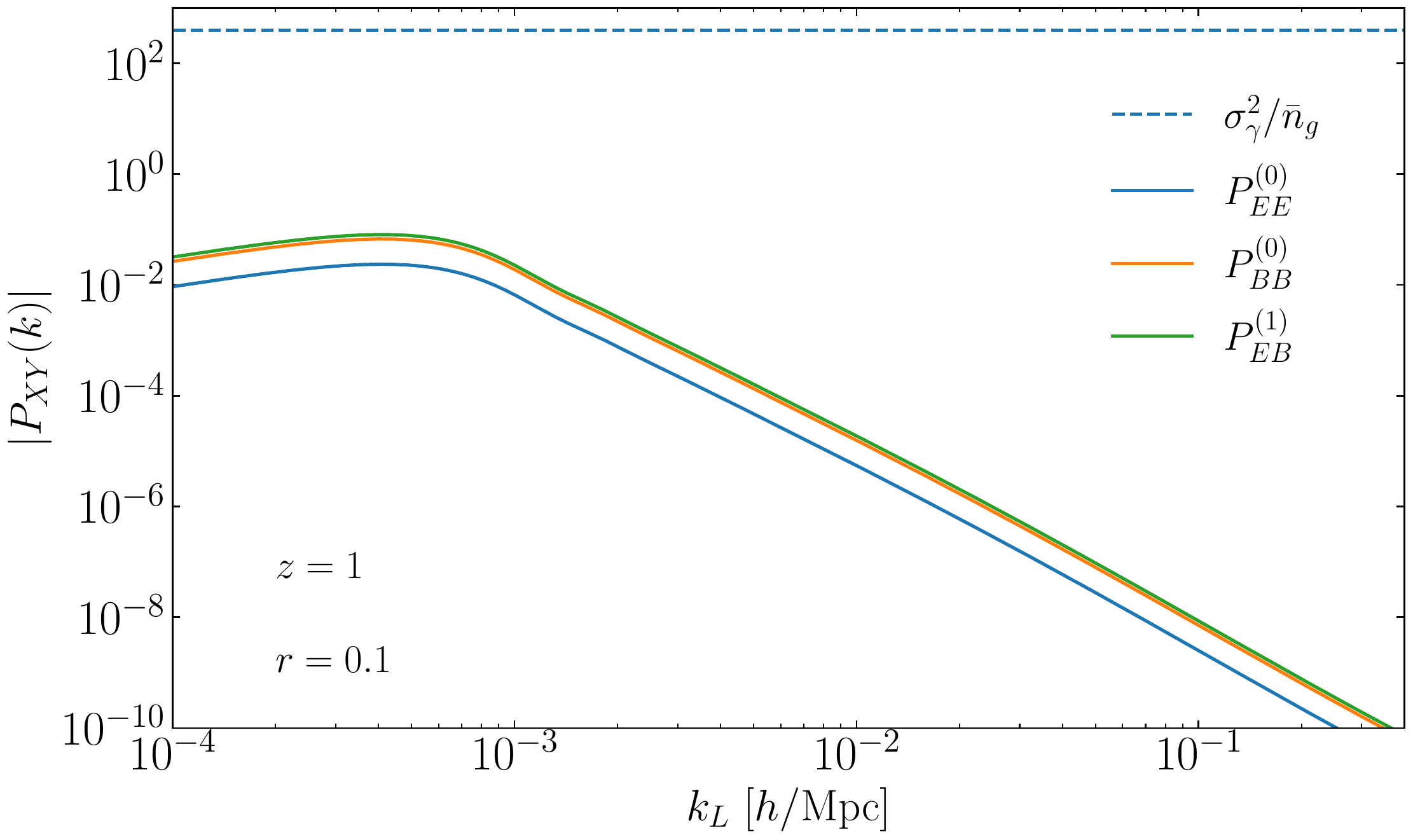}
\caption{
Shape auto-power spectra from GWs at $z=1$ assuming $r=0.1$.
The blue, orange, and green lines show the monopole of the $E$-mode auto-, the monopole of the $B$-mode auto-, and the dipole of the $EB$ cross-power spectra, respectively.
For the $EB$ cross power spectrum we also assume a maximally parity-violating case, i.e., $\chi(k)=1$. 
The blue dashed line depicts the shape noise.
}
\label{fig:shape_power}
\end{figure}

Let us discuss possible observables that could be used to probe GWs from LSS.
As GWs affect halo shapes at linear order shown in Sec.~\ref{sec:IA_GW}, the simplest probe would be the shape correlation, namely the intrinsic alignment from GWs~\cite{Schmidt:2012nw,Schmidt:2013gwa,Biagetti:2020lpx}.
Although the shape correlation is primarily affected by the scalar perturbations, we can use the $E/B$-decomposition to distinguish the GWs contribution from the scalar contribution at linear order.
Under the flat-sky approximation and assuming the line-of-sight direction, $\hat{n}$, is parallel to the $z$-axis, we can define $E$-mode and $B$-mode via
\begin{align}
    E({\bf k},\hat{n}) \pm i B({\bf k},\hat{n}) \equiv {_{\pm2}\gamma}({\bf k},\hat{n})e^{\mp 2i\phi_k},
\end{align}
where
\begin{align}
    {_{\pm2}\gamma}({\bf k},\hat{n}) \equiv m_\mp^i(\hat{n}) m_\mp^j (\hat{n}) \gamma_{ij}({\bf k}),
\end{align}
with ${\bf m}_{\pm}\equiv (1, \mp i, 0)/\sqrt{2}$.
At linear order, the scalar perturbations only induce $E$-mode with vanishing $B$-mode~\cite{Hirata:2004gc}.
On the other hand, GWs generate both $E$-mode and $B$-mode as
\begin{align}
    E({\bf k}_L,\hat{n}) &= b_K^{\rm GW}(k_L) \frac{1}{8}(1+\mu_L^2)\sum_{\lambda} h_{(\lambda)}({\bf k}_L),
    \\
    B({\bf k}_L,\hat{n}) &= - b_K^{\rm GW}(k_L) \frac{i}{2}\mu_L \sum_{\lambda} \frac{\lambda}{2}h_{(\lambda)}({\bf k}_L),
\end{align}
where we have used Eq.~\eqref{eq:IA_GW} and $\mu_L \equiv \hat{k}_L\cdot\hat{n}$.
Note that here we project the halo shapes onto two-dimensional plane but do not project their position; we can combine photometric and spectroscopic surveys to get the projected shapes and their three-dimensional positions~\cite[e.g.,][]{Okumura:2020,Kurita:2020hap,Akitsu:2020jvx,Kurita:2022agh}.
Defining the power spectra for $E$- and $B$-modes via
\begin{align}
    \langle X({\bf k})Y^*({\bf k}') \rangle \equiv (2\pi)^3 \delta^{(3)}_{\rm D}({\bf k}-{\bf k}') P_{XY}({\bf k}),
\end{align}
we obtain
\begin{align}
    P_{EE}(k_L, \mu_L;z) &= \frac{1}{64}(1+ \mu_L^2)^2 (b_K^{\rm GW}(z))^2 P_h(k_L) \simeq  \frac{49}{1024}(1+ \mu_L^2)^2\alpha^2(k_L;z) (b_K^{\rm scalar}(z))^2 P_h(k_L),
    \\
    P_{BB}(k_L, \mu_L;z) &= \frac14 \mu_L^2 (b_K^{\rm GW}(z))^2 P_h(k_L) \simeq  \frac{49}{64}\mu_L^2\alpha^2(k_L;z) (b_K^{\rm scalar}(z))^2 P_h(k_L),
\end{align}
and $P_{EB} = 0$ for unpolarized GWs, where we have used Eq.~\eqref{eq:bK_ansatz}.
Considering the chiral GWs, there appears a non-vanishing $EB$ correlation as
\begin{align}
    P_{EB}(k_L, \mu;z) &= \frac{i}{16}\mu_L(1+\mu_L^2) (b_K^{\rm GW}(z))^2  \chi(k_L) P_h(k_L)
     \simeq i\frac{49}{256} \mu_L(1+\mu_L^2)\alpha^2(k_L;z)  (b_K^{\rm scalar}(z))^2 \chi(k_L)P_h(k_L),
\end{align}
where $\chi$ is defined in Eq.~(\ref{eq:chi}). 
Notice that $P_h(k_L)$ is the primordial power spectrum of the tensor mode.

In Fig.~\ref{fig:shape_power}, we plot the lowest order moment of the multipoles of these spectra at $z=1$, i.e., the monopole of the $EE$ and $BB$ spectra and the dipole of the $EB$ spectrum, assuming $b_K^{\rm scalar}=0.1$, $r=0.1$, and $\chi(k)=1$ as a demonstration.
On large scales, the suppression comes from the behaviour of $\alpha(k_L)$ while on small scales the shape obeys a power-law with $P(k_L)\propto k_L^{-3}$ because $\alpha(k_L)$ does not change much.
We also display the shape noise as the dashed line, assuming $\sigma^2_\gamma=0.2$ and $\bar{n}_g = 5.0\times10^{-4}~(h/{\rm Mpc})^3$.
As is evident, for all cases the shape noise contribution is much greater than the expected signals, meaning that the detection of GWs using these spectra is very challenging.
There should also be the contributions of the scalar-mode both in $EE$ (at linear order) and $BB$ (at one-loop order),
though they are absent in the $EB$ spectrum~\cite{Biagetti:2020lpx}.
\footnote{Ref.~\cite{Biagetti:2020lpx} argued that the shape noise is absent in the $EB$ power spectrum, making it a cleaner probe of the chiral GWs.
Although this is true at the signal level, the covariance of the $EB$ spectrum includes the $EE$ and $BB$ spectra with the shape noise.
Thus even for the $EB$ power spectrum the detectability is limited by the shape noise.
}
Still, this sort of probes allows us to put the upper limit on the amplitude of GWs that are generated after the recombination or the reionization, which is not constrained by the CMB.
Specifically, we could obtain the 1-$\sigma$ error on $r$ of order $\sigma(r)\sim 10^3$ at $k\sim 10^{-3}~h/{\rm Mpc}$, implying the total energy spectrum of GWs, $\Omega_{\rm GW}(k)$, could be constrained as $\Omega_{\rm GW}(k) \lesssim 10^{-8}$ at $k\sim 10^{-3}~h/{\rm Mpc}$ from the current galaxy surveys.

The density-density-shape bispectrum also involves the GWs contribution.
For instance, the tree-level bispectrum of $\delta_{\rm h}$-$\delta_{\rm h}$-$B$ in squeezed limit ($k_L\to 0$) is found to be
\begin{align}
    \lim_{k_L\to 0}iB^{\rm grav.}_{{\rm hh}B}(k, k_L;z) =& 
    \frac12 b_1(z)  b_K^{\rm GW}(k_L;z) 
    \left[\left(2b_1(z)\alpha(k_L;z) + 2b^{\rm GW}_{s^2}(k_L;z) + b_1(z)\beta(k_L;z)\frac{\partial}{\partial\ln{k}}   \right)P_{\rm lin}(k;z)\right]
    \nonumber\\
    &\hspace{8.5cm}\times\mu_L \hat{k}^i\hat{k}^j\sum_\lambda \frac{\lambda}{2}e^{(\lambda)}_{ij} (\hat{k}_L) P_{h_{(\lambda)}}(k_L)
     \\
    \simeq &
    \frac78 b_1(z) b_K^{\rm scalar}(z) \alpha(k_L;z)
    \left[\left(2(b_1(z) + \frac74 b^{\rm scalar}_{s^2}(z))\alpha(k_L;z) + b_1(z)\beta(k_L;z)\frac{\partial}{\partial\ln{k}}   \right)P_{\rm lin}(k;z)\right]
    \nonumber\\
    &\hspace{8.5cm}\times\mu_L\hat{k}^i\hat{k}^j\sum_\lambda \frac{\lambda}{2}e^{(\lambda)}_{ij}(\hat{k}_L) P_{h_{(\lambda)}}(k_L),
\end{align}
where we neglect the contributions from the projection effect~\cite{Schmidt:2012ne} and the redshift-space distortion~\cite{Kaiser:1987qv}.
Notice again that here $P_{h_{(\lambda)}}(k_L)$ is the primordial power spectrum and all the redshift dependence is encoded in the bias coefficients ($b_K^{\rm GW}(k_L;z)$ and $b^{\rm GW}_{s^2}(k_L;z)$), $\alpha(k_L;z)$, and $\beta(k_L;z)$.
The superscript ``grav.'' emphasises that this bispectrum is induced by the gravitational interaction in the late time universe. 
In other words, there could be an additional contribution from the primordial universe, e.g., the scalar-scalar-tensor non-Gaussianity~\cite{Endlich:2012pz,Domenech:2017kno}.
In principle we can directly observe the scalar-scalar-tensor non-Gaussianity by looking at this density-density-shape bispectrum.

Another promising observable to probe GWs from LSS is the quadrupolar anisotropic imprint in the local density power spectrum discussed in Sec.~\ref{sec:power_response} and Sec.~\ref{sec:tidal_bias}.
One can construct the optimal quadratic estimator for GWs by using the anisotropic imprint in the local power spectrum~\cite{Masui:2010cz,Jeong:2012df,Dimastrogiovanni:2014ina,Dai:2013kra,Masui:2017fzw}.
Our result can be used to increase $k_{\rm max}$ of the estimator in this method, allowing for the improved detectability.
As we demonstrated for the first time how GWs affect the biased tracer, a more realistic estimator for the biased tracer can be available.
Taking the cross correlation with the CMB would also be valuable to improve the detectability~\cite{Dodelson:2010qu,Alizadeh:2012vy,Chisari:2014xia,Philcox:2022dht}.
We leave the detailed investigations on these possibilities for future work.

Let us conclude by mentioning one potentially significant effect of GWs on LSS observables.
In this paper, we focused on the effect of GWs on the cold dark matter (CDM) perturbations and ignored the effect on the photon-baryon fluid in the early universe.
However, given that the biased tracers such as galaxies and halos trace the CDM-baryon perturbation,
this could leave a distinctive signature in LSS observables as well.
We will investigate it in future work.

\begin{acknowledgments}
We thank Giovanni Cabass, William Coulton, and Matias Zaldarriaga for insightful discussions and Fabian Schmidt for valuable comments on the draft.
KA is supported by JSPS Overseas Research Fellowships. 
TO acknowledges support from the Ministry of
Science and Technology of Taiwan under Grants Nos. MOST
110-2112-M-001-045- and 111-2112-M-001-061- and the Career Development Award,
Academia Sinina (AS-CDA-108-M02) for the period of 2019 to 2023.
Numerical computation was carried out on the Helios and the Typhon cluster at the Institute for Advanced Study and Popeye-Simons cluster at San Diego Supercomputer Center.
The Flatiron Institute is supported by the Simons Foundation.
\end{acknowledgments}

\appendix
\section{From global FLRW coordinates to local CFC coordinates}
\label{app:CFC}
Here we outline the mapping from the global perturbed FLRW metric to the local CFC coordinates. More detailed discussion can be found in Refs.~\cite{Pajer:2013ana,Schmidt:2013gwa,Dai:2015rda}.
We start from the perturbed FLRW metric,
\begin{align}
    ds^2 = a^2(\eta)\left[-(1 + h_{00}(\eta,{\bf x}))d\eta^2 +\left(\deltaK_{ij} + h_{ij}(\eta, {\bf x}) \right)dx^i dx^j  \right],
\end{align}
where we neglect $h_{0i}$ component.
The coordinate transformation is given by
\begin{align}
    x^0 &= x_F^0 + \frac12 \int_0^{x^0_F} h_{00}(\tilde\eta) \dd \tilde\eta 
    +v_i x_F^i
    - \frac14 h'_{ij}x_F^ix_F^j,
    \\
    x^i& = x_F^i + v^i(x^0_F - \eta_F) - \frac12 h^i_{\ j}x_F^j 
    - \frac14\left[ h^i_{\ j,k} + h^i_{\ k,j} - h_{jk}^{\ \ ,i}\right]x_F^jx_F^k,
\end{align}
where $v_i$ is the coordinate velocity of the central geodesic and $h_{\mu\nu}$ is evaluated along the central geodesics.
The CFC metric can be obtained by using the transformation law of the metric, $g_{\mu\nu}^F(\eta_F,  {\bf x}_F) = g_{\alpha\beta}(\eta, {\bf x})(\partial x^\alpha/\partial x^\mu_F)(\partial x^\beta/\partial x^\nu_F)$, yielding
\begin{align}
    g_{00}^F &= -a_F^2(\eta_F)\left[ 1 - \frac12 \left( h_{00,ij}+ \cH h'_{ij} + h''_{ij} \right)x_F^ix_F^j \right],
    \label{eq:g00_CFC}
    \\
    a_F(x^0_F) &= a\left(\eta = \eta_F + \frac12 \int_0^{\eta_F} h_{00}(\tilde\eta)\dd \tilde\eta\right).
\end{align}
In Eq.~\eqref{eq:g00_CFC},
setting $h_{00}=0$ coincides to the main text (Eqs.~\eqref{eq:CFC_metric}-\eqref{eq:CFC_tides}).
On the other hand, the case where $h_{00} = -2 \Phi_L$ and  $h_{ij} = 0$ corresponds to the usual tidal separate universe simulation picture (see App.~\ref{app:scalar_tidal_bias}).

\section{The tidal responses to the scalar tides from tidal separete universe simulations}
\label{app:scalar_tidal_bias}

\begin{figure}[tb]
\centering
\includegraphics[width=0.6\textwidth]{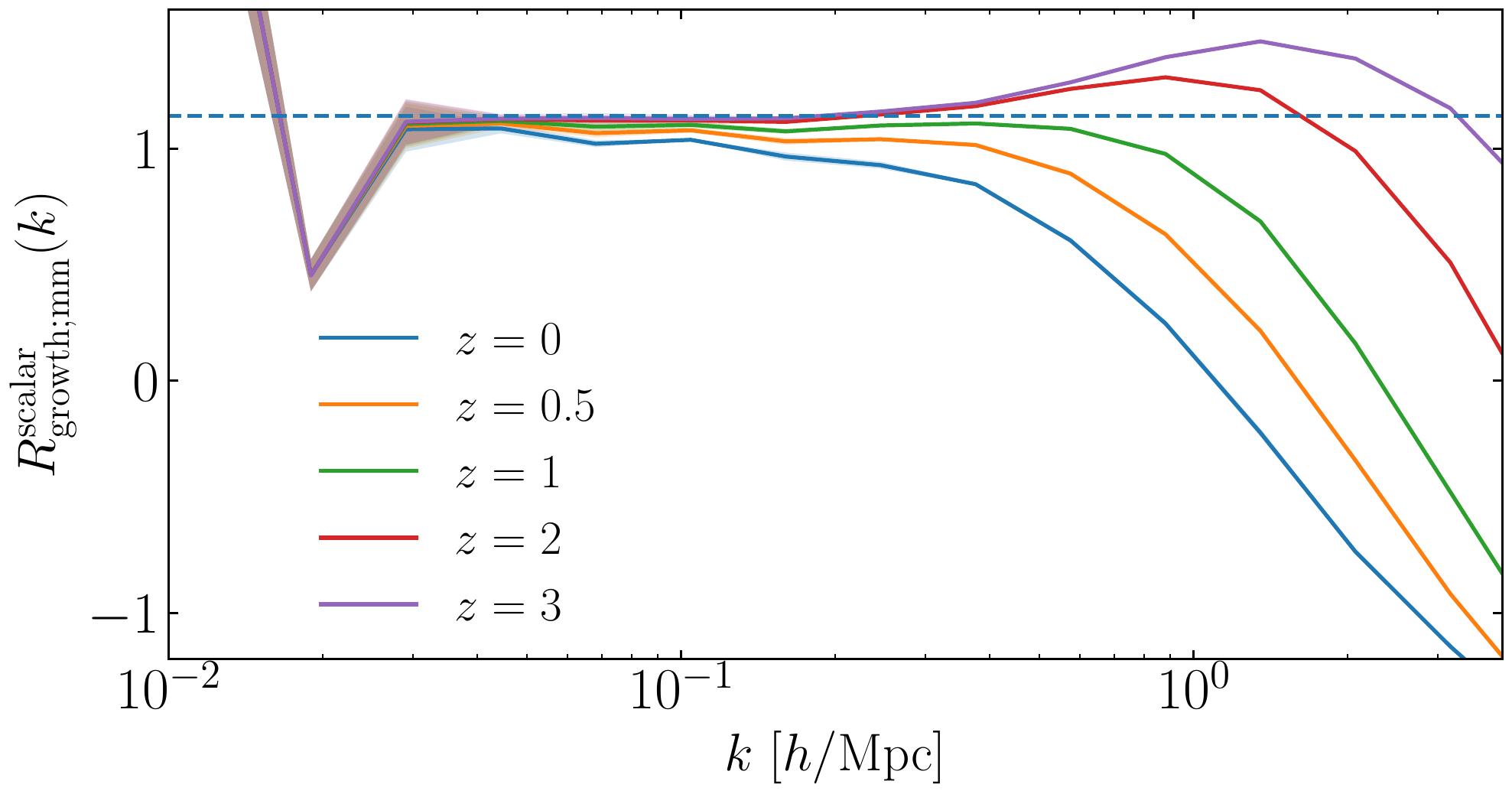}
\caption{
Growth response of matter auto-power spectrum to the scalar tides, $R^{\rm scalar;mm}_{\rm growth}(k)$ for various redshifts, measured from the simulations.
The blue dashed line corresponds to the perturbation theory prediction, $R^{\rm scalar}_{\rm growth;mm} = 8/7$~\cite{Akitsu:2016leq}.
}
\label{fig:Rk_mm_scalar}
\end{figure}

In this appendix, we summarize the tidal responses to the scalar tidal field.
These include not only the tidal response of the matter auto-power spectrum and the linear shape bias, which are already presented in Refs.~\cite{Akitsu:2020fpg,Stucker:2020fhk}, but also the tidal responses of halo-matter and halo auto-power spectra and the halo tidal bias.

\begin{figure}[tb]
\centering
\includegraphics[width=\textwidth]{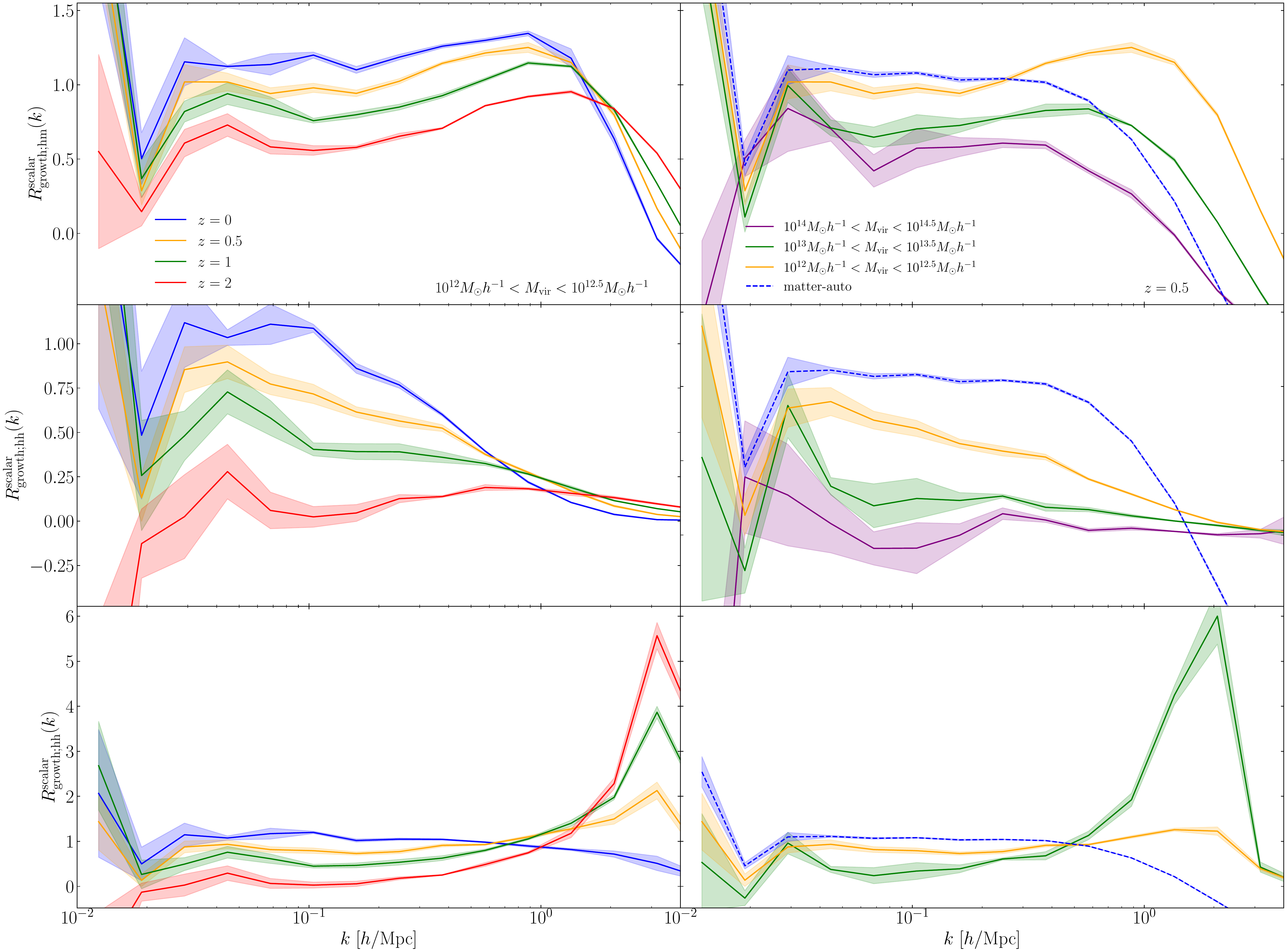}
\caption{
Growth responses involving the halo density field as a function of $k$.
In the left panels the responses for various resfhits are plotted while in the right panels the responses for various halo masses are plotted.
The top panels show the response of the halo-matter cross-power spectrum, $R^{\rm scalar}_{\rm hm;growth}(k)$,
The middle panels show the response of the halo auto-power spectrum, $R^{\rm scalar}_{\rm hh;growth}(k)$, normalized by the halo auto-power spectrum with the shot noise, and the bottom panels show the response of the halo auto-power spectrum, $R^{\rm scalar}_{\rm hh;growth}(k)$, normalized by the halo auto-power spectrum without the shot noise.
}
\label{fig:Rk_hm_hh_scalar}
\end{figure}

First, we sketch the construction of the tidal separate universe in the scalar tides case, focusing on the difference to the GWs case, although we refer the reader to Ref.~\cite{Akitsu:2020fpg} for details. 
The difference between the GWs case and scalar-tides case in the induced tidal field in the local region, $\tau_{ij}$, results in the different initial conditions and anisotropic scale factors.
In the scalar-tides case, the induced tides become
\begin{align}
    \tau_{ij}(\eta) = -\frac12 h_{00,ij} &=  \Phi_{L,ij}
    \\
    & =  \frac12 \Omega_{\rm m}(\eta) \cH^2 \delta_L(\eta)\deltaK_{ij} + \frac32 \Omega_{\rm m}(\eta) \cH^2 s_{L,ij}(\eta),
    \label{eq:CFC_tides_scalar}
\end{align}
where we have used Poisson equation and decomposed into the trace (large-scale overdensity, $\delta_L$) and the traceless (large-scale pure tidal field, $s_{L,ij}$).
The construction of the tidal separate universe with the scalar tides corresponds to replacing Eq.~\eqref{eq:CFC_tides} with Eq.~\eqref{eq:CFC_tides_scalar} and repeating the analysis in Sec.~\ref{subsec:2LPT_GWs}, Sec.~\ref{subsec:TSU_background}, and Sec.~\ref{subsec:TSU_ICs}.
Focusing on the pure tidal mode ($s_{L,ij}$), the equation that governs the evolution of the anisotropic scale factors, $\Delta_i$, is now
\begin{align}
    \Delta''_i(\eta) + \cH\Delta_i'(\eta) 
    = -\frac32 \Omega_{\rm m}(\eta)\cH^2 s_{L,i}(\eta),
\end{align}
whose solution is given by
\begin{align}
    \Delta_i(\eta) = -s_{L,i}(\eta_0) \frac{D(\eta)}{D(\eta_0)}.
    \label{eq:Delta_i_scalar}
\end{align}
The perturbative prediction of the growth response is
\begin{align}
    \lim_{k \to 0}R_{\rm mm;growth}^{\rm scalar}(k) &= \frac87,
    \label{eq:Rk_scalar_pert}
\end{align}
where the response with respect to the scalar tides is defined through
\begin{align}
    \left.\frac{\dd{\ln P_{\cal G}}}{\dd{s_{L,ij}}}\right|_{{\bf k}_{\cal G}} =
    \left.\frac{\dd{\ln P_{\cal L}}}{\dd{s_{L,ij}}}\right|_{{\bf k}_{\cal G}}
    =&
    \left.\frac{\partial\ln P_{\cal L}}{\partial s_{L,ij}}\right|_{{\bf k}_{\cal L}}
    +\left.\frac{\partial\ln P_{\cal L}}{\partial \ln k_{{\cal L},\ell}}\right|_{s_{L,ij}}
    \left.\frac{\dd{\ln k_{{\cal L},\ell}}}{\partial s_{L,ij}} \right|_{{\bf k}_{\cal G}}
    \nonumber\\
    \equiv&\,
    \hat{k}_i\hat{k}_j
    \left[
    R^{\rm scalar}_{\rm growth}(k)
    +R^{\rm scalar}_{\rm dilation}(k)
    \right],
    \label{eq:def_response_scalar}
\end{align}
which is analogue to Eq.~\eqref{eq:def_response_GW}.
One important difference of Eq.~\eqref{eq:def_response_scalar} from Eq.~\eqref{eq:def_response_GW} is that in Eq.~\eqref{eq:def_response_scalar} we define the response with respect to the scalar tides at the \textit{same} epoch rather than the initail epoch.
As a result, the perturbative prediction remains the same for all redshifts.
Also note that these results are independent of the wavenumber of long-modes.

Fig.~\ref{fig:Rk_mm_scalar} shows the tidal response of the matter auto-power spectrum to the scalar tidal field as a function of $k$ for various redshifts, measured from tidal separate universe simulation with the scalar tides.
This should be contrasted with Fig.~\ref{fig:Rk_gw_kLs_redshifts} in the main text.
In Fig.~\ref{fig:Rk_hm_hh_scalar} we show the tidal responses of the halo-matter cross-power spectrum (the top panels), the halo auto-power spectrum with and without the short noise (the middle and the bottom panels respectively) for various redshifts (the left panels) and for various halo masses (the right panels).
This is analogue to Figs.~\ref{fig:Rk_gw_hm}-\ref{fig:Rk_gw_hh} in the main text.
The differences of the responses for various redshifts and halo masses on large scales should be explained by the halo biases (the combination of $b_1(z;M)$ and $b^{\rm scalar}_{s^2}(z;M)$).
The peaky feature appeared in the halo auto response normalized by the halo auto-power without the shot noise is due to the non-Poissonian feature of the shot noise term and thus not physical (see the discussion in the last paragraph in Sec.~\ref{subsec:growth_halo_GW}).

\begin{figure}[tb]
\centering
\includegraphics[width=\textwidth]{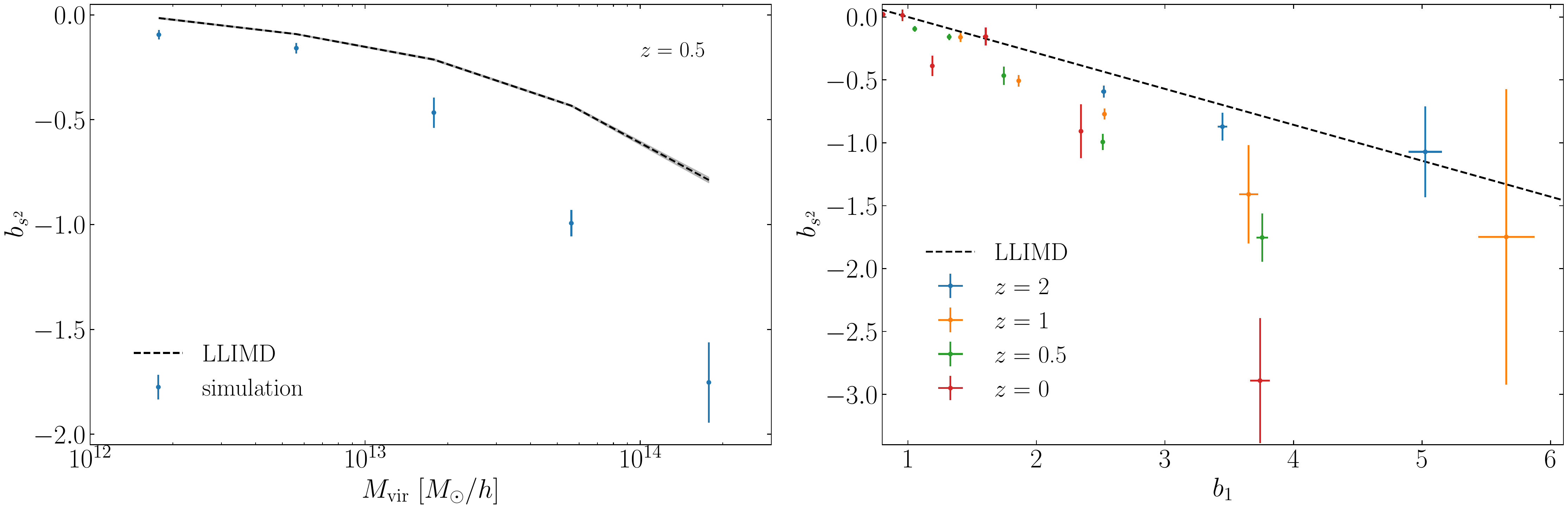}
\caption{
Halo tidal bias measured from the halo-matter power spectrum response. \textit{The left:} The tidal bias as a function of halo mass at $z=0.5$. 
\textit{The right:} The tidal bias as a function of the linear bias $b_1$ from various redshifts and halo masses.
The black dashed line displays the Lagrangian local-in-matter-density (LLIMD) prediction: $b_{s^2}^{\rm scalar} = -\frac27 (b_1 - 1)$.
}
\label{fig:bs2_scalar}
\end{figure}

\begin{figure}[tb]
\centering
\includegraphics[width=0.6\textwidth]{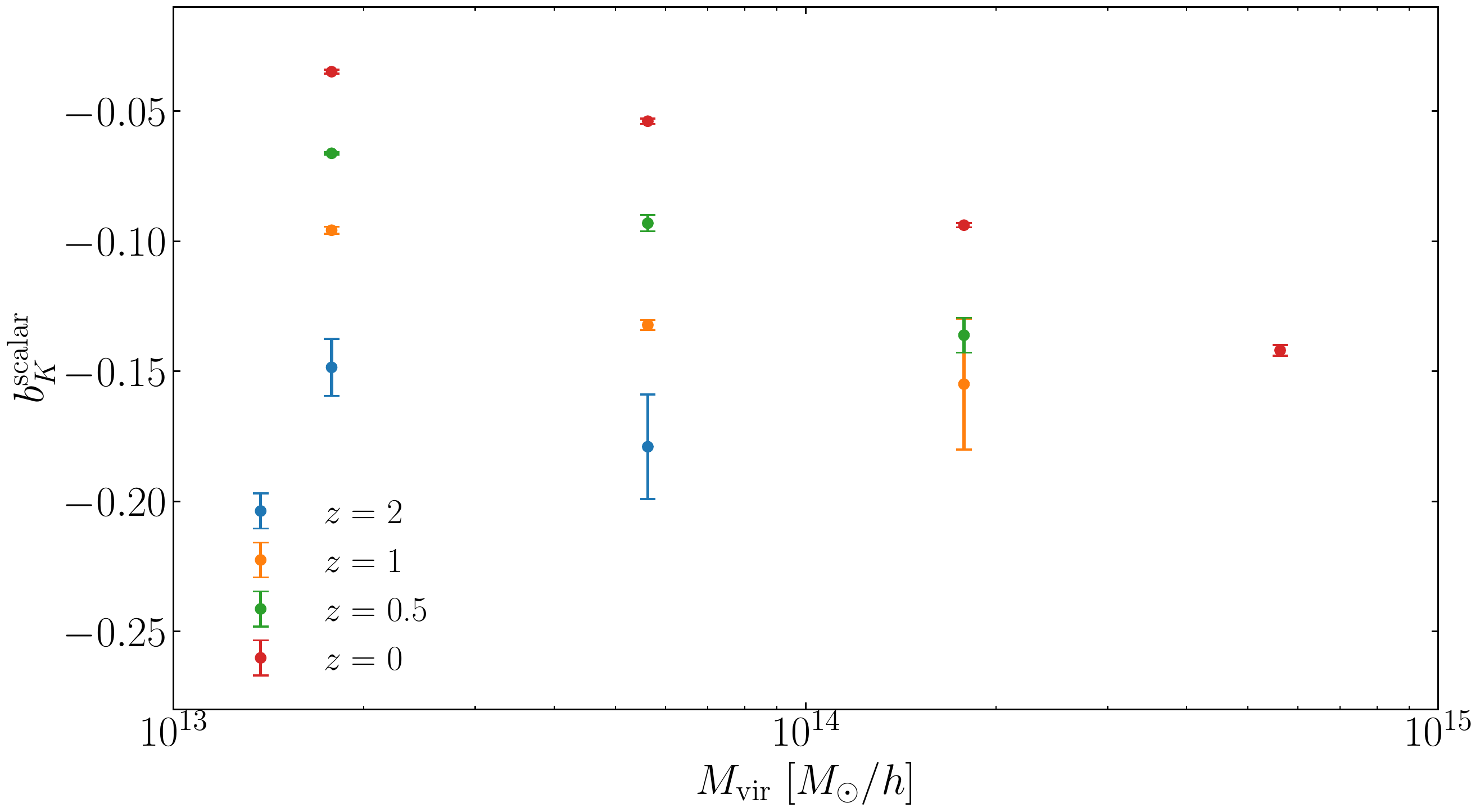}
\caption{
Linear shape bias induced by the scalar tides as a function of halo mass for various redshifts.
}
\label{fig:bK_scalar}
\end{figure}

We measure the halo tidal bias $b^{\rm scalar}_{s^2}$ in the same way as Eq.~\eqref{eq:local_b1_GW} and Eq.~\eqref{eq:bs2_GW_estimator} in the main text.
The result is presented in Fig.~\ref{fig:bs2_scalar} where the left panel shows $b_{s^2}^{\rm scalar}$ at $z=0.5$ for various halo masses and the right panel shows $b_{s^2}^{\rm scalar}$ as a function of $b_1$.
We compare the result with the Lagrangian local-in-matter-density (LLIMD) prediction~\cite{Desjacques:2016bnm}: $b_{s^2}^{\rm scalar} = -\frac27 (b_1 - 1)$, which is plotted as the black-dashed line.
In general the LLIMD prediction fails to capture the behaivour of $b_{s^2}^{\rm scalar}$, in particular at high-mass end, which is consistent with Refs.~\cite{Lazeyras:2017hxw,Abidi:2018eyd}.

In Fig.~\ref{fig:bK_scalar} we show the linear shape bias (or the linear alignment coefficient) induced by the scalar tides, $b_K^{\rm scalar}$, (introduced in Eq.~\eqref{eq:IA_scalar}) as a function of halo mass for various redshifts.
The estimator for $b_K^{\rm scalar}$ used here is the same as Eq.~\eqref{eq:bK_estimator}.
Fig.~\ref{fig:bK_scalar} is a counterpart of Fig.~\ref{fig:bK_GW_mass}.

\section{The modification in the drift operator}
\label{app:drift}

\begin{figure}[tb]
\centering
\includegraphics[width=0.48\textwidth]{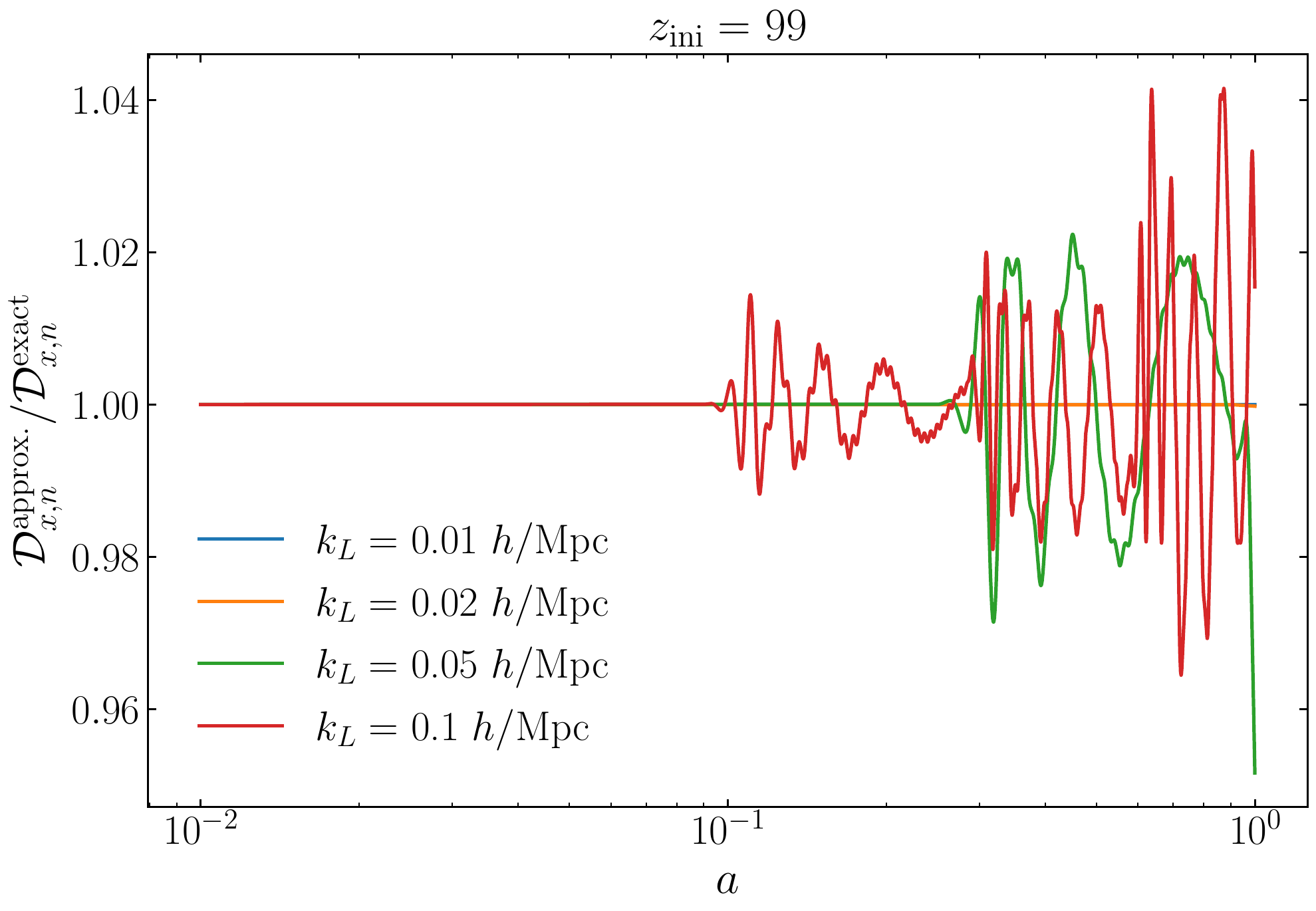}
\hfill
\includegraphics[width=0.49\textwidth]{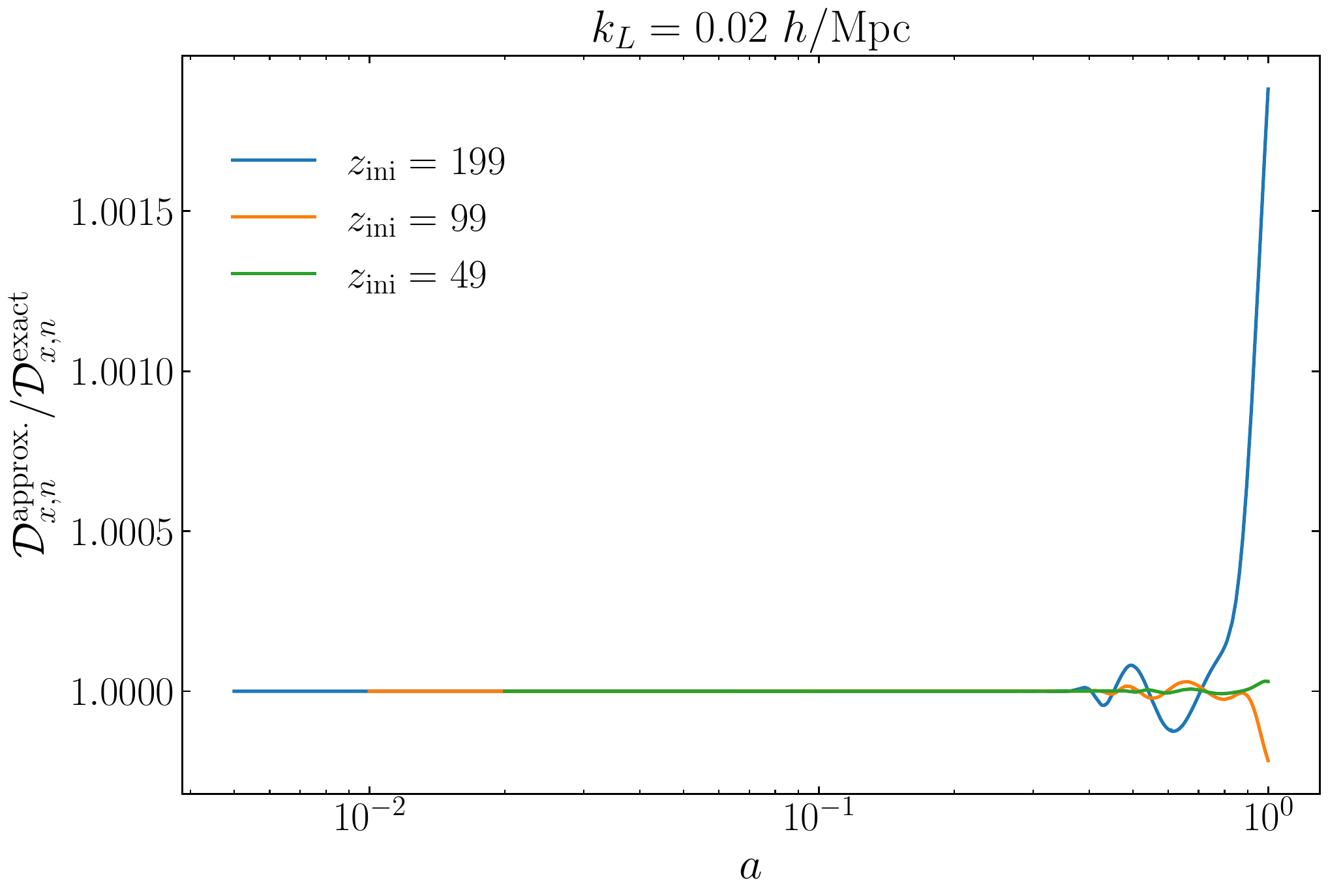}
\caption{
Ratio of the approximate drift integral to the exact one as a function of the scale factor.
\textit{Left:} the ratio for different wavenumber of GWs with the same starting redshift $z_{\rm ini}=99$. 
\textit{Right:} the ratio for different starting redshifts with the same wavenumber of GWs $k_L=0.02~h/{\rm Mpc}$. 
}
\label{fig:drift}
\end{figure}

In the tidal separate universe simulation, the drift operator changes from the usual $N$-body simulation as 
\begin{align}
    x_{i;n+1} = x_{i;n} + \frac{P_i}{m} \int^{a_{n+1}}_{a_n} \frac{\dd a}{a_i^2 \cH}
    \equiv x_{i;n} + \frac{P_i}{m}~ {\cal D}^{\rm exact}_{i;n} ,
    \label{eq:drift}
\end{align}
where $P_i$ is the conjugate momenta of $x_i$, $i=x,y,z$ and $n$ represents the time step.
In the original \texttt{Gadget-2} code~\cite{Springel:2005mi},  instead of computing the drift integral in Eq.~\eqref{eq:drift} at each time step, first it prepares the following table
\begin{align}
    {\cal D}_i[j] \equiv \int^{a[j]}_{a_{\rm ini}} \frac{\dd a}{a_i^2 \cH},
\end{align}
where $1\leq j \leq N_{\rm drift}$ with $N_{\rm drift}$ being the length of the table.
$a[j]$ is the $j$-th scale factor that is sampled equally spaced in logarithm from $a_{\rm ini}$ to $a=1$, regardless of the actual time step.
The actual drift integral at each time step is then evaluated by linearly interpolating this drift table as
\begin{align}
    {\cal D}^{\rm approx.}_{i;n}
     = &
    \int^{a_{n+1}}_{a_{\rm ini}} \frac{\dd a}{a_i^2 \cH} 
     - \int^{a_{n}}_{a_{\rm ini}} \frac{\dd a}{a_i^2 \cH} 
     \nonumber
     \\
    \simeq& \ 
    {\cal D}_i[j+1] + \left(a_{n+1} - a[j+1]\right) \left({\cal D}_i[j+2] - {\cal D}_i[j+1]\right)
    \nonumber
    \\
    &- {\cal D}_i[j] - \left(a_{n} - a[j]\right) \left({\cal D}_i[j+1] - {\cal D}_i[j]\right)
\end{align}
with $j=\lfloor n \rfloor$.

This prescription works well for a monotonic integrand, which is the case for the usual cosmological simulation.
We found, however, that this approximation for the drift operator breaks down for the tidal separate universe simulation with GWs where the integrand oscillates.
Fig.~\ref{fig:drift} compares the drift factor evaluated by the above procedure and the direct calculation of the integral in Eq.~\eqref{eq:drift}.
Although the default length of the drift table is 1000 we increased it to $N_{\rm drift} = 200000$.
Even with this large table, the approximated drift factor fails to capture the exact result.
Therefore we modified the drift operator so that at each step the drift integral is directly evaluated without using the drift table or interpolation.
This modification is particularly important for $k_L \gtrsim 0.02~h/{\rm Mpc}$ when $z_{\rm ini} = 99$.
In fact, without this modification the results suffers from the artifact.

\bibliography{references}
\end{document}